\documentclass[aps,prd,showpacs,nofootinbib,floats,floatfix,reprint,preprintnumbers,groupedaddress,onecolumn]{revtex4-2}
\usepackage[normalem]{ulem} 
\usepackage{bm}
\usepackage[titletoc]{appendix}
\usepackage{latexsym}
\usepackage{dcolumn}
\usepackage{amsmath,amsfonts,amssymb}
\usepackage{graphicx,epsfig}
\usepackage{amsmath}
\usepackage{fancyhdr}
\usepackage{hyperref}
\usepackage{graphicx,epstopdf}
\usepackage{xcolor}
\hypersetup{
	colorlinks   = true, 
	urlcolor     = blue, 
	linkcolor    = blue, 
	citecolor   = red 
}

\usepackage[T1]{fontenc}

\def\lsim{\lower.5ex\hbox{$\; \buildrel < \over \sim \;$}}
\def\gsim{\lower.5ex\hbox{$\; \buildrel > \over \sim \;$}}
\def\simeq{\lower.3ex\hbox{$\; \buildrel \sim \over - \;$}}

\def\ie{{\it i.e.,} }
\def\mk{\dot{M}_{\rm K}}
\def\mt{\dot{M}_{\rm T}}
\def\mkc{\dot{M}_{\rm Kc}}
\def\ma{\dot{M}_{\rm A}}
\def\mab{\dot{M}_{\rm AB}}
\def\med{\dot{M}_{\rm Edd}}

\def\rt{R_{\rm t}}
\def\rs{R_{\rm g}}
\def\as{a_{\rm s}}

\begin{document}

\title{A new prospect for jet origin and spectral state transitions with three hybrid accretion flows around black holes}
%
%
\author{Rajiv Kumar}\email{rkumar@ustc.edu.cn}
\author{Ye-Fei Yuan}\email{yfyuan@ustc.edu.cn}
\affiliation{CAS Key Laboratory for Research in Galaxies and Cosmology, Department of Astronomy, University of Science \& Technology of China, Hefei,
Anhui 230026, China}
\affiliation{School of Astronomy and Space Sciences, University of Science \& Technology of China, Hefei 230026, China}
\date{\today}

\begin{abstract}
There is a very popular two-zone accretion disk model that the inner part of the non-advective Keplerian disk (Shakura-Sunyaev disk) can produce hot advection-dominated accretion flow, which can generate high energy power-law radiation and outflows/jets. However, we find that this simple model is inadequate to automatically explain many properties of the sources (such as hysteresis effect, counter-clockwise traversal in a hardness-intensity diagram, peculiar variabilities, and association/non-association of jets in a black hole $X-$ray binary) without considering additional assumptions. We also find some theoretical issues in this model, such as understanding of variation of transition region and formation of an outer disk, which has only Keplerian distribution. Based on the recent theoretical studies on advective disk structures, as well as, many observational behaviors of the accreting black holes, we conclude that there should be a third component (TC) of accretion flow parallel to the two-zone disk model, which can naturally explain all above mentioned issues.
Interestingly, this modified model also provides a new scenario for the jet generation and evolution with the TC flow during high energy states, which can make the jet close to the axis. We also find out an expression of jet kinetic power. 
\end{abstract}
\maketitle
\section{Introduction} \label{sec:intro}
The black hole $X-$ray binaries (BXBs) are observed with many types of spectral states, and their transitions, which can be represented on the hardness-intensity diagram (HID) \cite{fbg04, rbm17, dzp21, drd21}. This HID is also called as q- diagram, as illustrated in the first plot of middle row in the Fig.\ref{fig1}. It is observed that a typical BXB follows the counter clockwise cycle in the HID from the quiescence state (QS) - low hard state (LHS) - high (hard/soft) intermediate states (HISs) - high soft state (HSS) - low intermediate states (LISs) to again reaches QS. Commonly, the bipolar jets are found in LHS to HISs \cite{mr98,fbg04}. When the jet components are separated out from the disk then the BXB enters HSS \cite[In case of full outburst][otherwise moved to other low luminous states]{drd21}. The QS is always found in the right side with lowest luminosity in the HID, which is believed to be a starting point of the outburst cycle in BXBs, and moved counter clockwise direction with rising luminosity. A soft state of a source is dominated with thermal multicolor blackbody (BB) emission, and a hard state is dominated with non-thermal power law emission. 
Moreover, the quasi-periodic oscillations (QPOs) in power-law component are also observed during LHS and HIS \cite{rm06}.

\begin{figure}
\begin{center}
\includegraphics[width=0.45\linewidth]{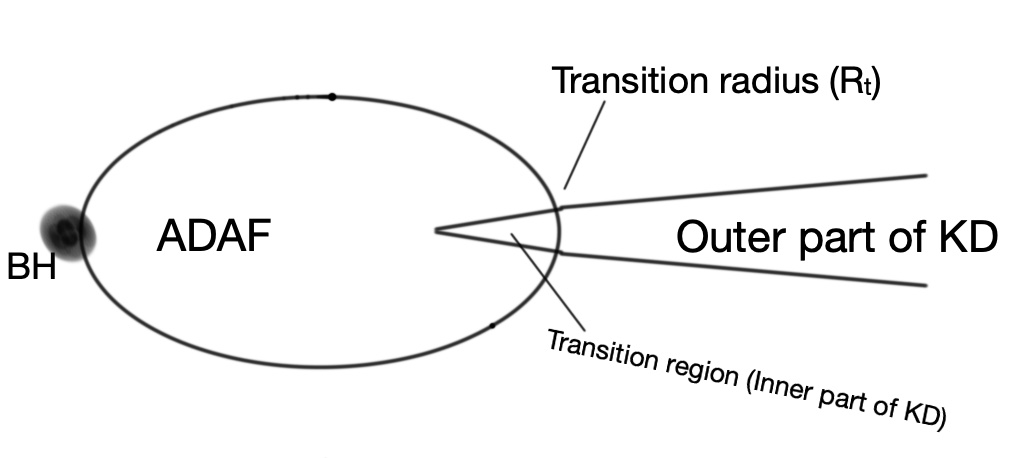}
\includegraphics[width=0.4\linewidth]{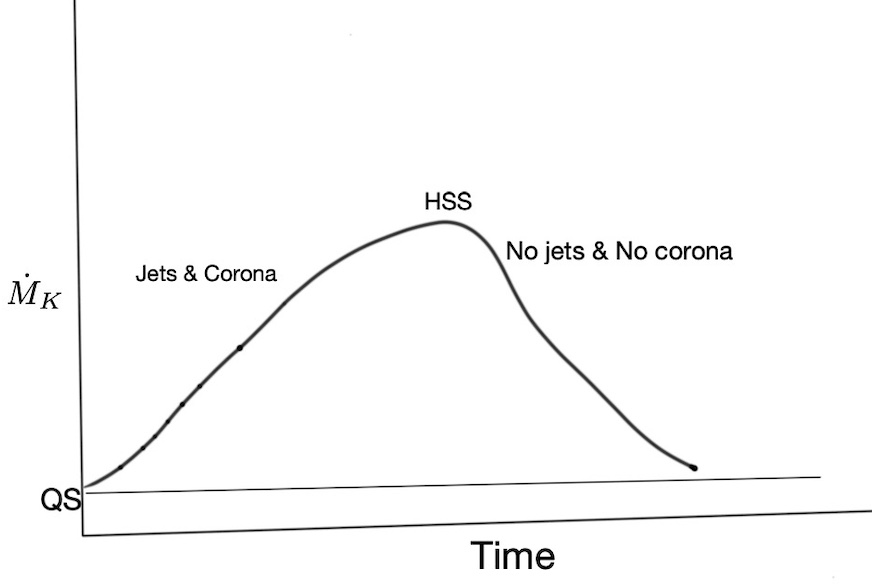}
\end{center}
 \caption{
Presented schematic diagrams of the two-zone disk model in left side, and variation of $\mk$ with a characteristic time scale of a cycle in the HID is presented in right-hand side. Here, $\mk$ is the accretion rate of the KD.
A rising part of the curve $\mk$ represents the right-path in HID, which has jet and corona. A decline part of the curve $\mk$ represents the left-path, which has no jet and no corona. The maxima of the curve can correspond to HSS in case of full outburst, or other low states in case of failed outburst.
 }
\label{fig00}
\end{figure}

The soft thermal or BB component in the spectral states is well explained by the Keplerian disk (KD)\citep{ss73}, which is optically thick and non-advective disk. A corona or hot gas (optically thin medium) is used for the generation of high energy power-law components in the studies of the accreting black holes (BHs). The origin and location of this corona is still under debate. For modeling purpose, some hot advective solutions  or lamppost model (unknown origin of the corona on the jet-axis or may be jet base) were proposed by the authors \citep{ct95,emn97,frs89}, which are used as the corona. Those advective solutions are hot and sub-Keplerian flow, which can also generate the outflows/jets \citep{ny95,yn14,lck16,gbc20,kc17,kg18}. In this regards, two types of the advective solutions in the accretion theory are widely used in the modeling studies, which are the shock solution \citep{c89}, and advection-dominated accretion flow (ADAF) solution \citep{ny94}. Interestingly, many authors have evidenced in their studies that the ADAF can be generated from the inner part of the KD due to some instabilities \citep[references therein]{le74,h96,emn97,kg19}. Thus the KD has two roles, one, a source of the BB radiation, and second, a gas source for generation of the ADAF in downstream side of the KD. So the whole disk has two different nature of flows or hybrid flows.
This hybrid flow theory is also known as two-component hybrid accretion flow model \citep{emn97}, and the coupling region of the KD and the ADAF disk is known as a transition region. Since both the inflows in this model are located on two different parts of the radial distances, so we can also call it hereafter the two-zone disk model (see schematic diagram in left-hand side of Figure \ref{fig00}).
Eventually, many studies have used this model to explain the spectral states and their transitions in the BXBs with adjusting the location of the transition region (can be denoted as a radius, $R_t$) and mass accretion rate \citep[][references therein]{emn97,fph06,mfp18}. This model is also used to explain the spectral properties of the active galactic nuclei (AGNs).
In BXBs, it is believed that the accretion rate increases during the QS - LHS - HISs - HSS path (hereafter, right-path) and decreases during HSS - LISs - QS path (hereafter, left-path) \citep{bmm11,kb15}. A schematic variation of accretion rate with characteristic timescale of a HID is represented in the right-hand side of Figure \ref{fig00}. 
However, both paths are different because of the presence of the hysteresis in the HID. The presence of the hysteresis is due to the presence of corona and jets in the right-path of HID.

The two-zone disk model consists of the KD flow as an outer part and the ADAF as an inner part of a disk around the BH. The KD flow can be further divided into two parts, one a transition region or an inner part of the KD, and other an outer part of the KD as shown in left-hand side of Figure \ref{fig00}.  An accretion rate of the ADAF (denoted as $\ma$, \ie basically depletion rate of the inner part of the KD) can be dependent on the accretion rate of the outer part of the KD (denoted as $\mk$) and size of the transition region. 
Generally, It is believed that the $\rt$ is moving towards BH with the increasing $\mk$ during the right path of the HID, and the $\rt$ is receding with the decreasing $\mk$ during the left path of the HID \citep{emn97,fph06,kb15}. 
So we expected that the overall average variation of $\mk$ should be as generic bell-shape form during a cycle of the HID as illustrated in right-hand side of Figure \ref{fig00}. 
In this model, the both components are always present in a whole cycle of HID with increasing or decreasing of the accretion rate, although the presence of each component can be stronger and weaker. 
So the two-zone disk model has same disk structures with the variation of $\rt$ in the whole HID. 
However, the HID has hysteresis and counter-clockwise traversal direction of the states. So this model seems not enough to explain the whole HID without addition of some proper accretion physics and processes, which can differentiate the paths of the HID and ensure the counter clockwise traversal direction.
Moreover, this model also has some additional theoretical issues. So the main points are following, 
\begin{enumerate}
\item 
An accreting BH can be feb by the various gas feeding mechanisms, which can have different nature of the inflow gases at the outermost accretion boundary (AB), such as hot gas ($T_{AB}\sim T_{vir}$), cold gas ($T_{AB}<<T_{vir}$), magnetized/non-magnetized gases, fully/partially ionized or neutral gases, and mixture of gases. 
The Keplerian angular momentum (AM) is a unique AM distribution in the  accretion disk, which maintains equilibrium between centrifugal force and gravitational force at each radii. So how do the viscous transportation of AM due to viscosity do consistently to maintain the Keplerian distribution in the large part of outer disk (means KD) in the two-zone model in spite of the BHs are fed by the different nature of inflow gases and feeding mechanisms? Moreover, the viscous stress tensor, $\tau_{\mu\nu}\propto T$ means AM transportation depends on the initial temperature of the inflow gases \citep[details in][]{ky21}. Here $T$ is a temperature of the inflow gas.
\item It is observed that the bipolar jet strength is increasing during the right-path of the HID.
However, we found that the global/local energy of generated ADAF decreases  with decreasing $\rt$ along the right path of HID (The details are presented in the next section). Furthermore, the local energy of whole ADAF becomes negative if $\rt\lsim$ a few $\times 10\rs$ \citep{kg19}, which makes stronger gravitationally bound inflow. So, it can require more extra energy (may be magnetic energy or thermal energy) to change the direction of the inflow to upward along the jet axis. 
According to the studies of \cite{ny95,bb04}, the positive energy of the accretion flow can help to make outflows/jets in the ADAF. So it is seemed that the jet generation from the ADAF in two-zone model is inefficient to make stronger jet with decreasing $\rt$. 
\item It is believed that the collapse\footnote{The collapse means the matter is quickly sucked by the BH when it crosses the ISCO radius due to no stable orbits inside the ISCO. Here the gravity pull becomes very strong and super-exceeds over all the other interactions/forces.} of hot ADAF flow when $\rt$ reaches the ISCO (Inner-most stable circular orbit), which triggers the disappearance of the jet from the disk, and it caused the sharp decrease in observed radio flux. So the state transition can happen from HIS to HSS in case of full outburst.  However, in case of failed outburst \citep[Restricted successful and Hard-only outbursts][]{drd21}, it is found that the HSS stage is not followed after the jet disappeared, and it means that a source still has significant hard $X$-ray component or hot ADAF flow in the two-zone disk model. So
why is jet disappeared from the disk without collapse of ADAF flow? %
This indicates that the jets are collapsed or generated by some additional factor or source, not just by the absence or presence of the ADAF. %
\item  It is generally accepted that $\rt$ is maximum in the QS and minimum in the HSS \citep[references therein]{bmm11,kb15}. So we believe that $\rt$ is approaching towards BH when $\mk$ increases in the right path, and receding when $\mk$ decreases during the left path of HID in the two-zone disk model. So the conditions for the variation of $\rt$ with following the conservation of mass flux should be, 
{\bf I- For approaching $\rt$:} $\ma<\mk$, which can allow to $\rt$ moves inward. 
{\bf II- For receding $\rt$:} $\ma>\mk$, which can allow to $\rt$ moves away from the BH, and 
{\bf III- For no change in $\rt$:} $\ma=\mk$, so $\rt$ has fix location. In all the situations, $\mk$ can be fixed or increased or decreased, which can depend on the variation of the accretion rate at the outermost AB. However, a variation of $\ma$ can mostly depend on the variation of size of the transition region (For instance, a smaller transition region in inner part of KD can produce ADAF with lower $\ma$).   
So what could be the physical process, which can allow $\ma$ is smaller or larger than $\mk$ in the transition region with allowing $\rt$ approaching or receding in the two-zone model? 
Since the outer part of the disk is always KD in nature with increasing/decreasing of accretion rate in the two-zone disk model. 
So, this model requires some extra/external physical process, which can control the relationship between $\mk$ and $\ma$. 
 \item There is also not much understanding about the relationship between variation of $\rt$ with $\mk$, such as dependence of $\rt$ on $\mk$ is same or different in both paths of the HID. If it is different then why is it so? For instance, \cite{kb15} has assumed totally different variations of $\rt$ with $\mk$ in both paths of HID. However, they have not given any theoretical explanation for such kind of $\rt$ variation with respect to $\mk$. 
\item Why has the jets generated in the right-path and not in the left-path of the HID in spite of the almost same disk structures are present in the whole HID with generic well-shape variation of the $\mk$ (See right side of Figure \ref{fig00}). 
\end{enumerate}
It seems that the above issues can not be naturally answered by the two-zone disk model without adding some extra assumptions or accretion process(es). For instance, \cite{kb15} had tried to answer the last two questions with considering some arbitrary assumptions, such as, the dependence of $\rt$ with $\mk$ is different in the both path for a typical HID, the jets can be generated with assumption of the presence of Poynting-Robertson cosmic battery during the right-path only. However, they have not mentioned the detail physics of those assumptions, such as, why does the relation between $\rt$ and $\mk$ is different in different states of the HID? Although, they have stated on page 3 that "Observationally, there is no agreement on the details of the variation of the transition radius as a function of accretion rate, but it is generally accepted that $\rt$ is large in the QS and in the LHS, when $\mk\lsim0.01\med$, and is at its minimum in the HSS." In their study, most of the explanations and other assumptions are based on the speculation of $\rt$ variation. They have also not mentioned in their study about the hysteresis in the failed outbursts, very high ISs and some other peculiar properties, which are varying much faster than the dynamical timescale of the KD, for instance, heartbeat states \cite[repeatedly moving between HSS and HIS][]{fbg04} in the GRS 1915+105, quick changes in the active galactic nuclei \cite[known as `changing-look' active galactic nuclei, CLAGN][]{pkk21,s21}.

Without proper answers of the above questions, It is difficult to understand the hysteresis effect in the HID with the two-zone disk model and some other observationed properties, such as,  counter clockwise traversal, very high IS, peculiar properties (shorter timescale extreme variabilities, heart-beat states, CLAGNs). Therefore, we have added some more physics into this model with addition of the one more accretion component, which is based on the recent studies of the general advective accretion theory \cite{ky21,kg19}.
We believe that a viable accretion model should be consistent with both theory and observations. Since we find that the two-zone model is largely inconsistent with both theory and observations, therefore, we are looking for the modifications in the model or new model of the accretion flow(s) around the BHs. In next section \ref{sec:GReqs}, we will briefly present the viscous relativistic fluid equations of motion and assumptions, section \ref{sec:resdes} is dedicated to present the accretion solutions and discussions of their properties, and final section \ref{sec:con} represents to concluding remarks and future improvements.
\section{Brief description of relativistic fluid equations and assumptions}\label{sec:GReqs}
We studied steady-state, axisymmetric, and relativistic viscous hydrodynamic accretion flows with vertical equilibrium around the Kerr BHs. Following \cite{ky21}, the conserve form of relativistic fluid equations can be written as
\begin{equation}
T^{\mu\nu}_{;\nu}=0~~\mbox{and}~~~(\rho u^\nu)_{;\nu}=0,~~~T^{\mu\nu}=(e+p)u^\mu u^\nu+pg^{\mu\nu}+\tau^{\mu\nu},
\label{ceq}
\end{equation}
where, $\rho$, $u$, $e$, $p$, $g^{\mu\nu}$, and $\tau^{\mu\nu}$ are local mass density, four-velocities, internal energy density \citep{cr09}, gas pressure, Kerr metric components, and viscous stress tensor, respectively. The indices ($\mu, \nu$) represent space-time coordinates and it is assumed in cylindrical coordinates ($t, r, \phi, z$).  
Taking disk assumptions, such as  axisymmetric ($\partial_\phi=0$), and vertical hydrostatic equilibrium in z-direction ($u_z\sim0$), and solved equations (\ref{ceq}) with assumed only $\tau^{r\phi}$ component of the viscous stress tensor is effective in the disk plane ($r-\phi$ plane). Then simplified form of above equations \cite[for details in][]{ky21} can be written as the bulk velocity gradient:
\begin{equation}
\frac{dv}{dr}=\frac{\frac{\as^2N}{N+1/2}\left[H_r+\frac{2}{r}-\frac{g_{rr,r}}{2g_{rr}} \right]+\frac{2\nu\sigma^{r\phi}\sigma_{r\phi}}{h(N+1/2)u^r}-{\cal F}}{\gamma_v^2\left[v-\frac{\as^2N}{v(N+1/2)}\right]},
\label{dveq}
\end{equation}
fluid temperature gradient:
\begin{equation}
\frac{d\Theta}{dr}=-\left[\Theta\left(\frac{u^r_{,r}}{u^r}+\frac{2}{r}+H_r\right)+\frac{\tilde{\tau}\nu}{u^r}2\sigma^{r\phi}\sigma_{r\phi}\right]/[N+1/2],
\label{dtheq}
\end{equation}
and angular momentum gradient:
\begin{equation}
\frac{dl}{dr}=-\frac{g_{rr}}{\gamma_v^2}\left[\frac{u^r}{\nu}(L-L_0)+\frac{1}{3}u^ru_\phi u^r_{,r}+\tilde{\sigma}\right],
\label{dleq}
\end{equation}
where  ${\cal F}=[1/r+\Delta^{'}/\Delta-D^{'}/D]/2$, $\Delta^{'}=2(r-1)$, $D=r^3+(r+2)a^2-(r-2)\lambda^2-4a\lambda$, and $D^{'}=3r^2+a^2-\lambda^2$,  $H_r=0.5/r+[2r(r^2+a^2)-\Delta^{'}a^2]/[(r^2+a^2)^2-2\Delta a^2]-[2r(r^2+a^2)+\Delta^{'}a^2]/[(r^2+a^2)^2+2\Delta a^2]$,  $u^r=\gamma_v v/\sqrt{g_{rr}}$,  $g_{rr}=r^2/\Delta$, $\Delta=r^2-2r+a^2$, $\gamma_v=1/\sqrt{1-v^2}$ is bulk Lorentz factor, $a$ is spin parameter of Kerr BH, $\lambda$ is a specific AM of the fluid and $v$ is bulk three-velocity of the fluid. 
The relativistic viscous stress tensor is $2\sigma^r_\phi =(g^{rr}+u^ru^r)u_{\phi,r}+\frac{1}{3}u^ru_\phi u^r_{,r}+\tilde{\sigma}$. Here $u_{\phi,r}$ and $u^r_{,r}$ are the ordinary derivatives of the four-velocities with respect to `$r$'. $\tilde{\sigma}=u^r_{;\phi}-g^{rr}\Gamma^\gamma_{\phi r}u_\gamma+u_\phi[u^r\Gamma^r_{r\gamma}u^\gamma+u^tu^r_{;t}+u^\phi u^r_{;\phi}]+u^r[u^t u_{\phi;t}-u^r\Gamma^\gamma_{\phi r}u_\gamma+u^\phi u_{\phi;\phi}]-2u^ru_\phi[u^t_{;t}+u^\phi_{;\phi}+u^z_{;z}+\Gamma^r_{r\gamma}u^\gamma]/3$, and $\Gamma^\gamma_{\mu\nu}$ is the Christoffel symbol. As we know that the BHs accretion is necessarily transonic in nature. However, the direct calculation of critical/sonic point is not possible with the above set of differential equations (\ref{dveq}-\ref{dleq}) because the equation (\ref{dleq}) has differential form, which is unlike to past studies with simple algebraic viscous prescriptions \citep{nt73,acg96,s09}. So we have developed a very sophisticated numerical scheme to find out sonic points and various types of transonic accretion solutions, and the detailed solution methodology is presented in \cite{ky21}. We have used five parameters in solving above model equations (\ref{dveq}-\ref{dleq}), which are grand specific energy parameter $E$, bulk angular momentum $L_0$ at the BH horizon, fluid composition parameter $\xi$, viscosity parameter $\alpha$, and BH spin parameter $a$.
\section{Results and discussions}\label{sec:resdes}
We have solved numerically the model fluid equations of motion (\ref{dveq}-\ref{dleq}) with using the two integration parameters ($E, L_0$), two fluid parameters ($\alpha, \xi$) and BH spin parameter ($a$). In present study, we have fixed $a=0.9$, $\alpha=0.01$ and $\xi=1.0$ in whole paper, and the integration parameters are changed for generation of the various accretion solutions. The values of integration parameters are directly/indirectly depended on the fuid parameters, BH spin parameter, and the outermost-accretion boundary conditions (OBCs). Interestingly, the OBCs can be suitably represented on the OBC-plane, which is a plane of the local energies of inflow gases at the outermost-accretion boundary locations (OBLs), and OBLs \citep[details in][corresponding Fig. 1 (Non-relativistic case), and Fig. 2a (Relativistic case), respectively]{kg19,ky21}. The OBC-plane has represented two types of inflow gases, which can generate all types of the general advective accretion flows. In present study, we have used three types of accretion solutions and their feasibilities, which are ADAF, smooth \citep[also named as ADAF-thick ][]{lgy99}, and shocked accretion solutions. The properties of these solutions are presented in following subsections.
\subsection{Properties of the ADAFs with $\rt$ variation}\label{sec:adaf}
\begin{figure}
\begin{center}
\includegraphics[angle=0, width=0.32\linewidth]{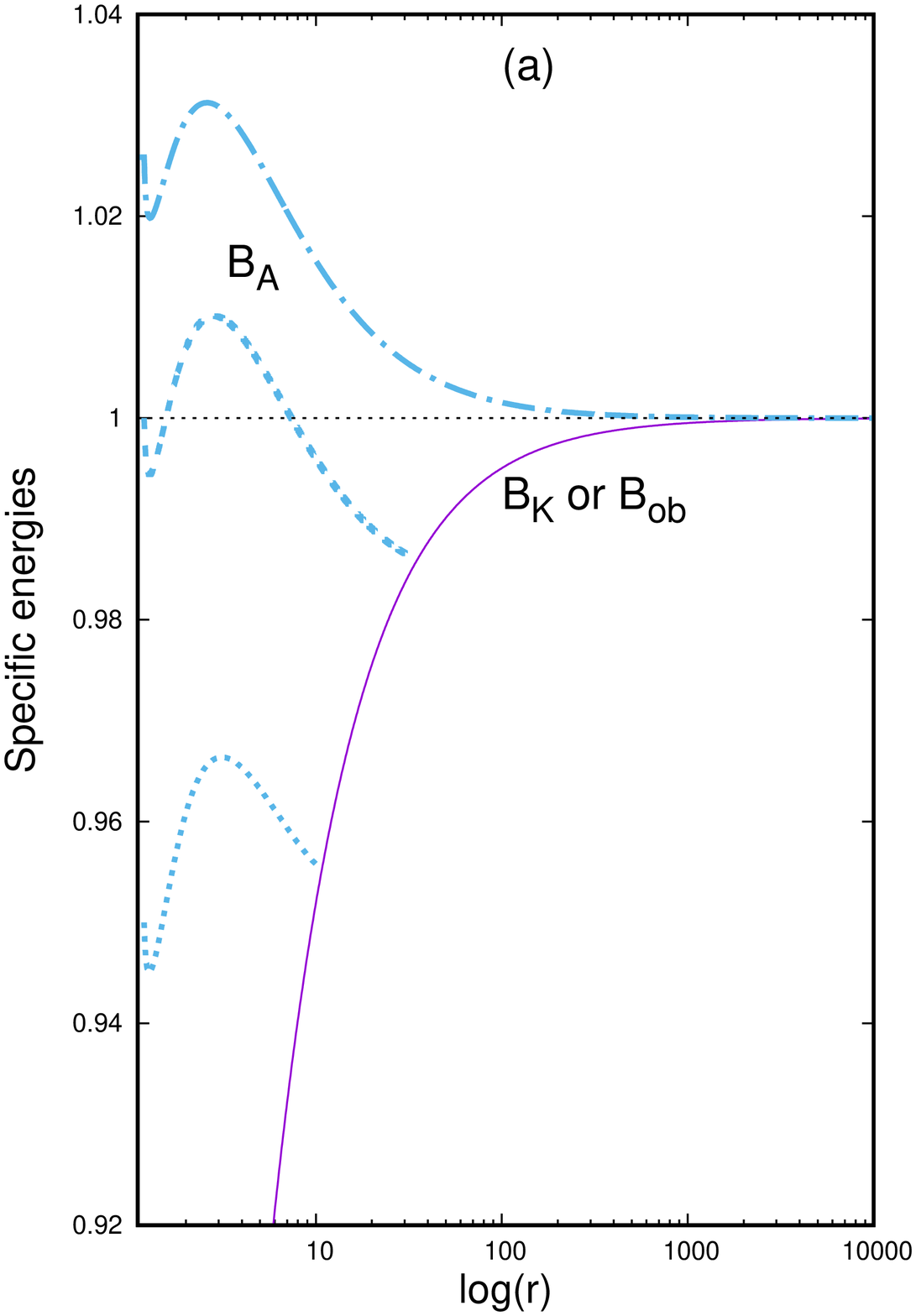}
\includegraphics[width=0.32\linewidth]{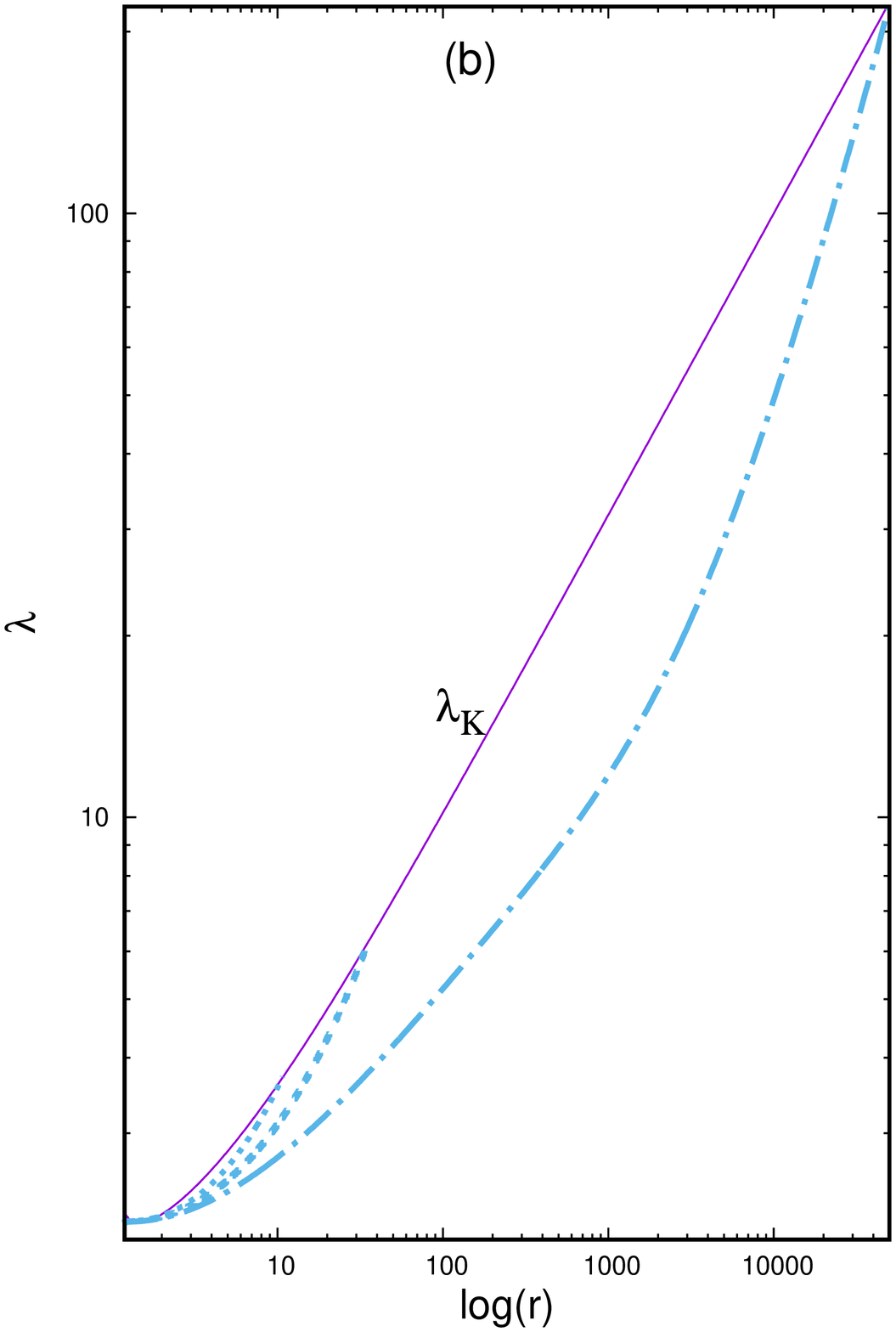}
\includegraphics[width=0.32\linewidth]{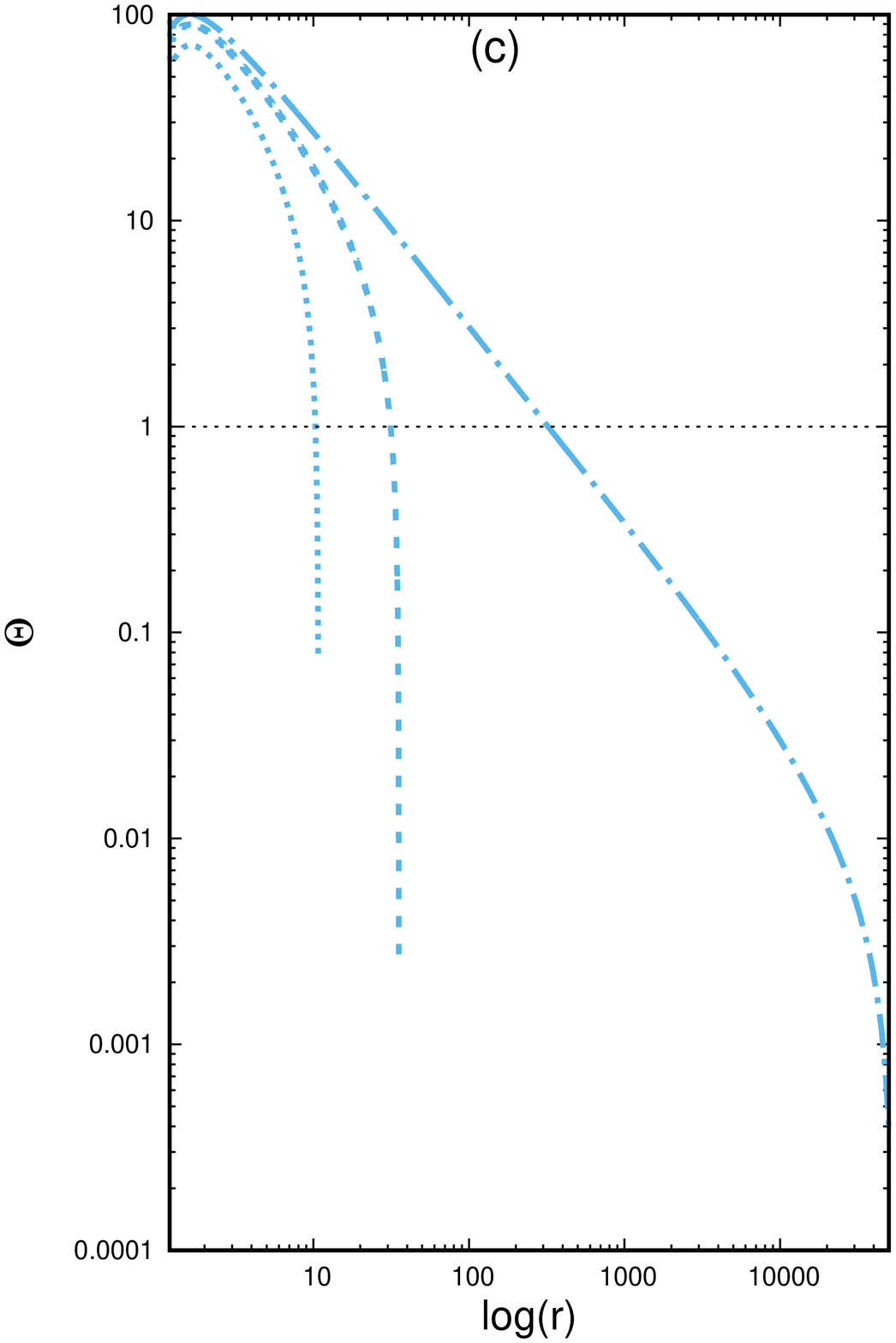}
\end{center}
\begin{center}
\includegraphics[width=0.32\linewidth]{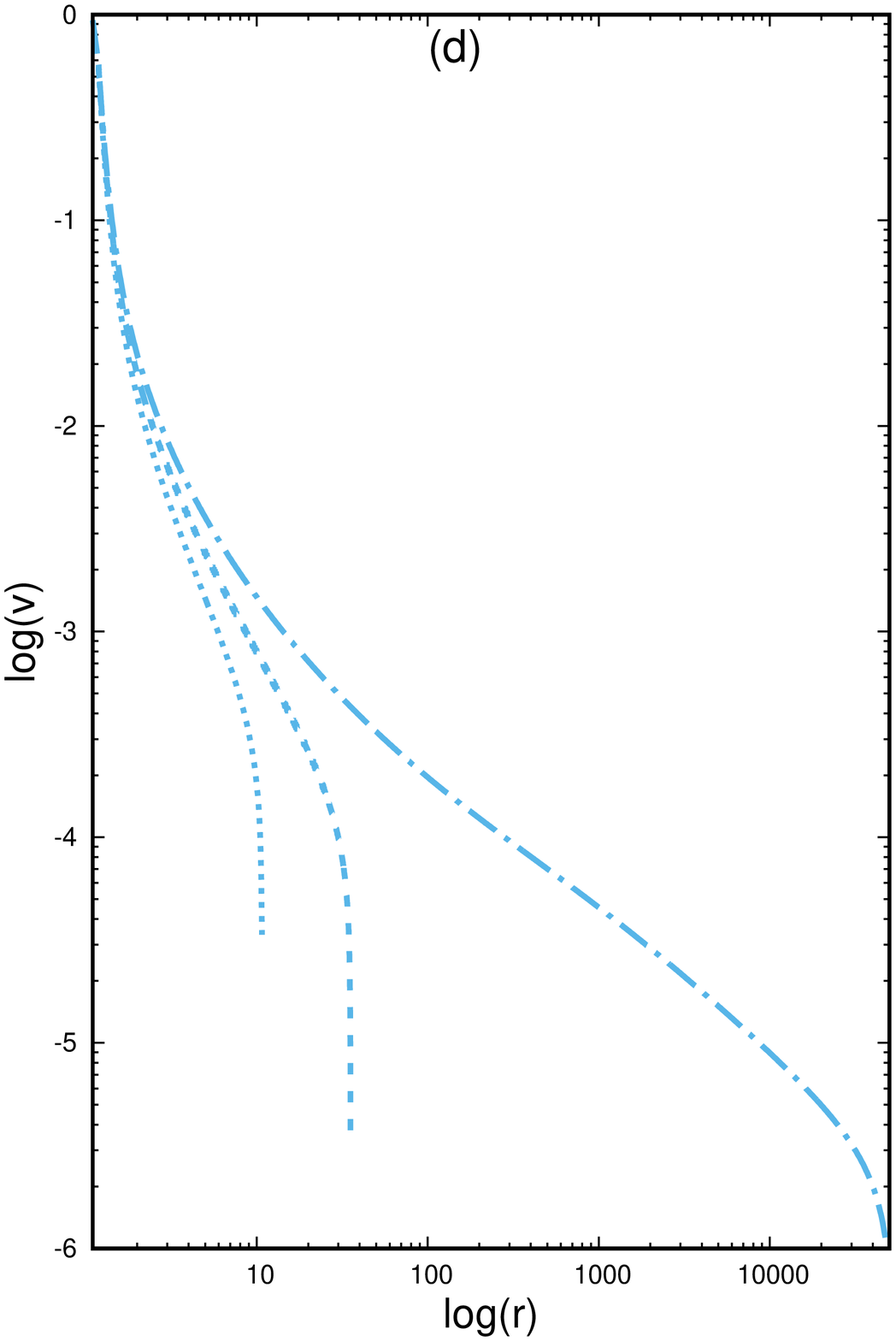}
\includegraphics[width=0.32\linewidth]{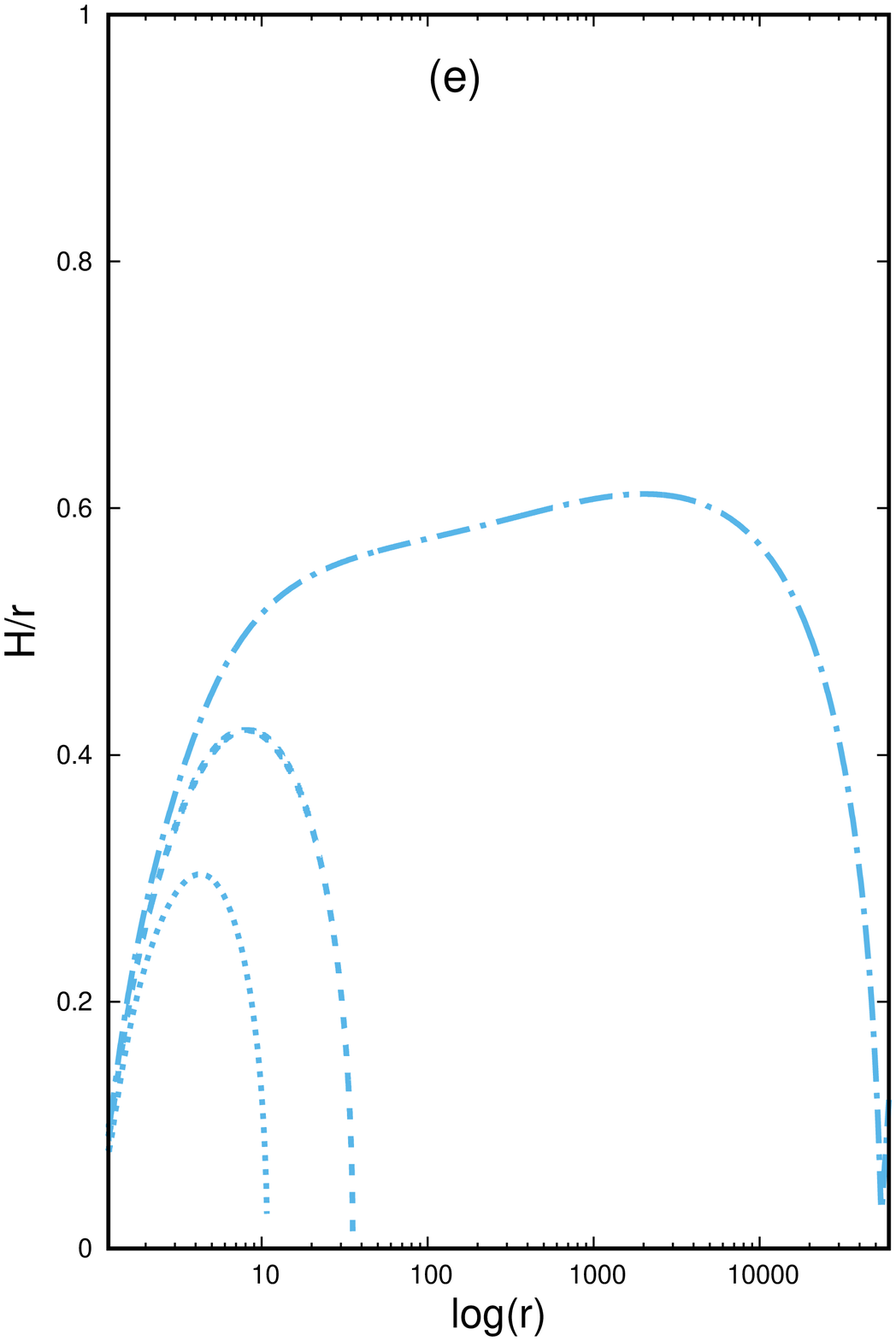}
\includegraphics[width=0.32\linewidth]{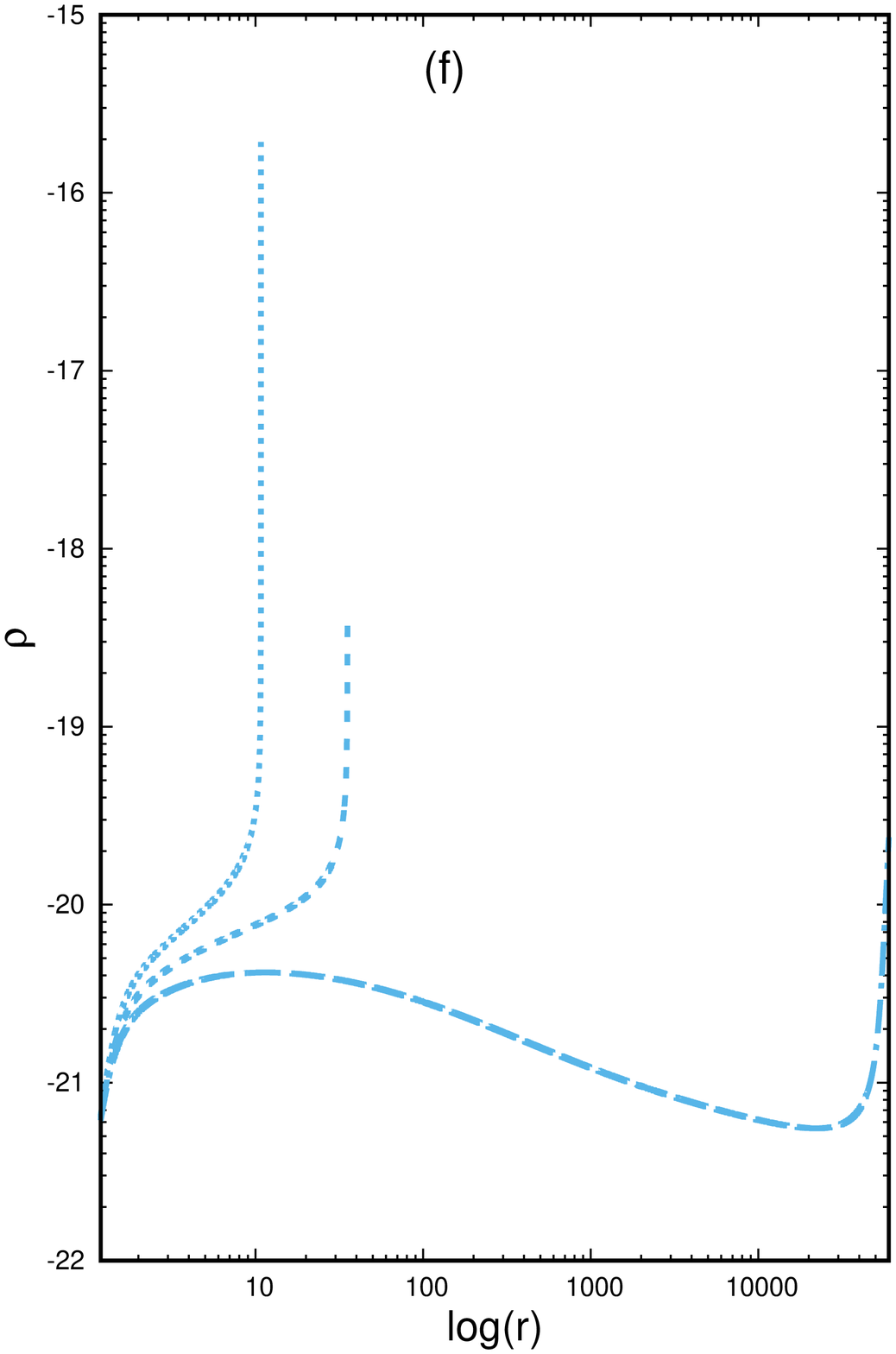}
\end{center}
 \caption{
Representation of the variations of the specific energies ($B_A$ and $B_K$, panel a), specific AM ($\lambda$, panel b), dimensionless temperature ($\Theta$, panel c), bulk inflow velocity ($v$, panel d), half-height disk ratio ($H/r$, panel e), and density ($\rho$, panel f) of the inflow gas with radial distance ($r$) are plotted. Here, $B_K$ and $\lambda_K$ are the local specific energy and specific Keplerian AM of the KD, respectively. These solutions are generated for the specific energy at outer-most AB, $B_{ob}=0.955613$ (dotted curve), $0.986139$ (dashed curve), and $0.999991$ (dotted-dashed curve). 
 }
\label{fig01}
\end{figure}

We believed that the ADAF can be generated from the inner part of the KD due to some instabilities, which can allow to evaporate the gas from the transition region of the KD \citep{le74,gl00,llg04,kg19}.  For comparison, we have plotted local energies of both the flows in Figure \ref{fig01}a, where $B_A$ and $B_K$ are the local relativistic Bernoulli energies for the ADAF and KD flow, respectively. Interestingly, $B_A$ at the outermost boundary of the ADAF is nicely matched with the local energy of the KD. Since it is believed that the location of the inner part (transition region) of the KD is changed with accretion rate in the whole HID. Therefore, a size of the ADAF can be changed with the variation of location of the transition region ($\rt$) of the KD. So here we have plotted the three different size of the ADAFs and their various flow variables in the panels. 
The rest of panels of the Figure \ref{fig01} are plotted with the flow variables of the three different size of the ADAFs, which are also supporting the scenario of generation of the ADAF from the KD flow, such as variation of $\lambda$ (panel b), $\Theta$ (panel c), $v$ (panel d), $H/r$ (panel e) and $\rho$ (panel f). Interestingly, all the basic flow variables of the ADAF at the outer AB are nicely matched or comparable to the flow variables of the inner transition region of the KD. So these things can provide strong support of generation of ADAF from the KD. Furthermore, $\rho$ at outer boundary of ADAF is higher, which can be comparable to the density of the KD. So, this kind of behavior of the $\rho$ can be 
 quite natural in the evaporation theory that the density should be higher around the source of evaporation and decreased away from the source till the evaporation process dominates over the accretion process. The accretion process around the central object is basically contraction of gas from large area to small area, which can increase the density of the inflow as in the middle part of the ADAF (panel f). 
 However, the $\rho$ of inflowing gas is again decreasing very close to the BH due to strong gravitational pull, which creates strong inflow velocity.
 Moreover, the outer boundary conditions (OBCs) for generation of the ADAFs are unique in comparison with the other types of advective solutions in the general accretion theory, for instance, $H/r<<1$ at AB for the ADAFs like KD flow, but other types of advective flows have $H/r\sim1$ at AB like quasi-spherical flow.  
 Theoretically, the ADAFs can not be obtained without considering viscosity in the flow, and other types of advective solutions can be obtained, e.g., shocked solutions, Bondi or Bondi-type, $\alpha-$types and smooth ADAF-thick solutions \cite{c89,smd22}. 

\begin{figure}
\begin{center}
\includegraphics[width=0.5\linewidth]{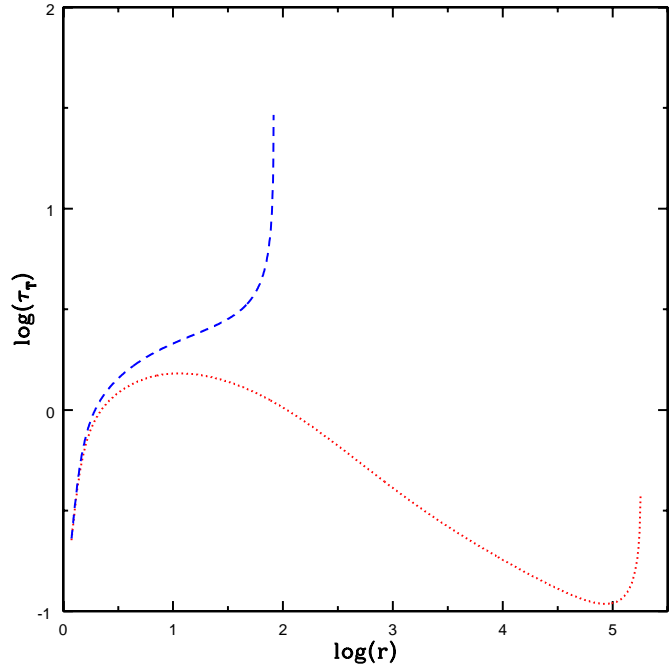}
\end{center}
 \caption{A typical behavior of the Thompson optical depth ($\tau_T=k_T\rho H$) for the two different size of the ADAFs. The accretion rate is used for both solutions, $\dot{M}=0.01\dot{M}_{Edd}$. 
 }
\label{fig3b}
\end{figure}

We have noticed some interesting properties with changing size of the ADAF due to changing $\rt$ in the two-zone disk model. Those properties can affect the nature of radiative emission and conditions of the outflow generation, such as positivity of the local energy in the accretion flow \citep{ny95,bb04}.  
The size of the accretion disk can also affect the total timescales of any radiative process and viscous processes, which can make overall affect on the total luminosity of a particular disk flow.  
The following interesting properties are
\begin{itemize}
\item In Figure \ref{fig01}a, the local specific energy $B_A$ of the ADAF decreases with decreasing $\rt$, and when $\rt<50\rs$ then the whole flow has $B_A<1$ (means negative energy). So the ADAF may not have outflows due to the negative local energy, which insists to keep gravitationally bound accretion flow. Moreover, the positivity of the local energy of the inflow is the necessary condition for the generation of the outflows \citep{ny95,bb04}. 
Observationally, it also found that the jet strength/kinetic energy increases during traversal of the right-path of the HID, which could be opposite, if the jet generation scenario from the ADAF due to decreasing energy of the ADAF with decreasing $\rt$. 
\item The AM distribution (Fig.\ref{fig01}b), and optical depth (Fig.\ref{fig3b}) of the ADAF is increased with the deceasing $\rt$. However, the temperature (Fig.\ref{fig01}c) and inflow velocity (Fig.\ref{fig01}d) is decreased with decreasing $\rt$. 
On the basis of variation of the density (Fig.\ref{fig01}f), the ADAFs can be divided into two types: One has decreasing and increasing density profile towards the BH, hereafter we can call it ADAF-I. And second has only decreasing density profile towards the BH, we can call it ADAF-II. 
The ADAF-II can be found when $\rt\lsim 500\rs$. For instance, the outer part of the ADAF-II can become optically slim (optical depth $\tau_T\gsim1$, for $\ma=0.01\dot{M}_{\rm Edd}$ as represented in the Fig.\ref{fig3b}), which can act as warm corona with $T<10^9K$ (See Fig.\ref{fig01}c). Furthermore, if $\rt<10\rs$ then the AM distribution of the flow becomes almost the Keplerian AM distribution with lower inflow velocity and lower temperature of the flow, which can behave like optically thick disk, and may contribute to the soft $X$-ray spectrum component or soft-excess.
\end{itemize}
So we can conclude that the small/intermediate size of the ADAF (or ADAF-II) can produce intermediate energy part of the spectrum (like soft-excess radiation), especially, the outer part of ADAFs with Compton thick scattering, which is close to the soft radiation source KD.
So the soft-excess (depends on accretion rate) can be suitably understood by the two-zone disk model. However, the ADAF-II may not generate the outflows or jets as the local energy of the flow is negative when $\rt<50\rs$.
The key point  is that the ADAF is changing their nature with changing size, so one should be very careful during modeling of the BXBs with adjusting the $\rt$. Moreover,  the qualitative power-law relation between $\rt$ and the flow variables of the ADAFs has been presented in \cite{kg19}, which can be used in the calculation of radiative emissivities and modeling of sources in the future studies.
\subsection{A disk with Three-component  flows}\label{sec:3cf}
{We know from the observed properties of the BXBs that the KD has weaker (larger $\rt$) or stronger (smaller $\rt$) presence allover the HID because of presence of the multicolor blackbody radiations in the spectrum of each spectral-state, so the two-zone disk model must be existed during the whole HID}. 
{Since the two-zone disk model is not sufficient to explain the many behaviors of the BXBs and AGNs as well as the theoretical points as discussed before. Therefore}, we believe that this model needs some extra physical process(es), which can ensure the counter clockwise traversal and hysteresis in the HID. 
The extra physical process(es) can be originated with addition of a third component (TC) of accretion flow along with the two-zone disk model. 
{This TC will be hot, optically thin and faster moving, which can also help to explain the high energy radiations (hard X-rays and $\gamma$-rays), jet generation, and some peculiar properties of the accreting BHs, such as, extreme variabilities with shorter timescales than the dynamical timescale of the KD}. So, the hybrid disk structure can have maximum three component of the accretion flow as details are in subsection \ref{sec:sum}. Furthermore, the presence of the TC flow should be dominated during the rising phase of outbursts of the BXBs means right-path of the HID. 
{From theory of the general accretion flow, we understood that the formation of the various advective solutions depend on the nature of the initial inflowing gases at the outermost AB, which changed the AM distribution of the flows due to nature of the viscous stress tensor ($\tau_{\mu\nu}\propto T$) \citep[details in][]{ky21}. The AM transportation due to viscous stress tensor is very sensitive to temperature of inflow gas around the outer AB, which can change the entire accretion flow nature. 
It is natural that the initial inflowing gases can have different temperatures, compositions and magnetization level. Moreover, the gas can be fully/partially ionized or neutral. All these inflow gases can affect the outward AM transportation due to the viscosity, so it can make different AM distribution in the disk, which can changed the flow properties and disk geometry. Recently, we have investigated the two-types of the initial inflowing gases, namely, hot-mode and cold-mode inflow gases, which give us the various kinds of the advective accretion flows due to their different AM distributions. So we believe that the accretion disk can have inflow gases with high (maximum $\lambda_K$) and low ($<\lambda_K$) AM distribution. Thus the possibility of the formation of only single AM distribution flow in the outer portion of the disk can be negligible, like, the only KD disk in the two-zone disk model. Since the AM of the flow has natural tendency to occupy region around the equatorial plane, therefore, the most natural configuration of the flows can be the high AM flow (like KD) on the equatorial plane and low AM flow (like TC) above the high AM flow, which can constituent the two parallel flows in the disk. So, this disk configuration can have total Three Hybrid Accretion flows (abbreviated as THAF) including the ADAF as presented in the Figure \ref{fig1}(panels B and C)}. Furthermore, the possibility of two parallel flow is much more in the outer portion of the disk than the other numbers of the flows because of the distinct two types of the inflow gases as found in the theoretical studies \citep{ky21,kg19}.  
The expected detail properties of the THAF will be explained in the following sections. Importantly, we would like to mention that the gas of KD can produce the ADAF/ADAF-like flows, which are mostly sub-sonic nature (becomes transonic very close to the BH), 
and the TC flow may undergo shock transition when the gas flow becomes super-sonic far away from the BH. 
So both advective flows (ADAF and TC) are having different nature and the TC is mostly hotter, lower AM and faster flow. Some expected properties of the THAF are presented in the following subsections.
\subsubsection{Momentum fluxes of the advective flows and Jet generation}\label{subsec:jet}
We calculated the momentum flux, $\dot{P}=\rho u^2+p$ of the accretion flows and represented in Figure \ref{fig3a}. Here, $\rho, u$ and $p$ are the mass density, bulk velocity and gas pressure of the flow, respectively. 
{The THAF model has the two parallel advective flows in the inner part of the disk, which are the ADAF and TC flow. So we would like to check the momentum inertia of the both flows, which can represent the existence of the both parallel flows without merging to each other in the disk. The momentum inertia can be understood in term of the momentum flux of the accretion flows. 
Interestingly, we find that the momentum flux is much higher for the ADAF (dashed blue line), and it also increases with the decreasing radius. So the ADAF can not be destroyed by the fast moving upper hot TC flow. Thus matter of the hot TC flow  can be deflected upward by the surface of higher momentum inertia ADAF disk. The deflected TC flow can move towards the rotation axis, which can form the spirally rotating jet around the axis. 
In the presented case, the $\dot{P}$ of the ADAF is almost two order of magnitude higher than the momentum flux of TC flow. So the surface of ADAF can effectively deflect the fast moving TC flow along the rotation axis. The strength or rigidity of the ADAF can be mostly dependent on the local specific AM and the momentum flux, which can make stronger deflection surface to deflect the TC flow.
We found in the study of accretion theory that the local specific AM distribution of the ADAF is always higher than the TC flow. The specific AM distribution and momentum flux of the ADAF is also increased with the decreasing  $\rt$, which can make stronger jet during traversal of the right-path of the HID.
\begin{figure}
\begin{center}
\includegraphics[width=0.5\linewidth]{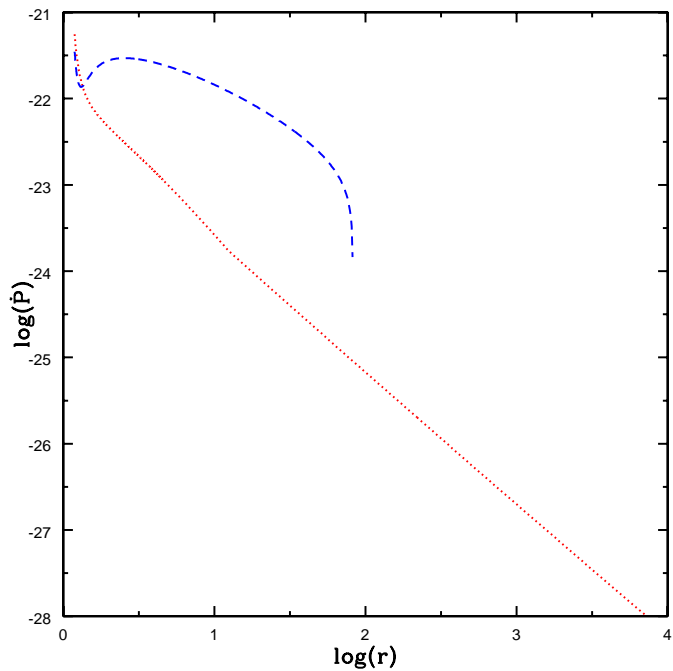}
\end{center}
 \caption{Variations of the momentum flux ($\dot{P}$) of the both parallel advective flows, upper TC flow (dotted red) and inner ADAF (dashed blue) is plotted in dimensionless geometric unit. Both flows are generated for $B_{ob}=0.993959$ (dashed curve) and $1.00001$ (dotted curve) with keeping same accretion rate, $\dot{M}=0.01\dot{M}_{Edd}$.
 }
\label{fig3a}
\end{figure}

Since the upper TC flow is participated in the jet formation in our THAF model, therefore the local energy flux of the TC flow can determine the kinetic power of the generated jet. In this scenario, an efficiency of the jet generation can depend on the local momentum fluxes of both parallel flows. Suppose, $\dot{P}_I$, and $\dot{P}_U$ are the local momentum fluxes of the ADAF, and upper TC flow at the interaction radius of both flows, respectively. Now, the efficiency of the jet emission can be defined as $\eta_j=(\dot{P}_I-\dot{P}_U)/\dot{P}_I$. The local energy flux of the TC flow is $\dot{B}_T=\dot{M}_T\bar{B}_T$, where $\dot{M}_T=\dot{m}_T\dot{M}_{Edd}$, and $\bar{B}_T=B_Tc^2$ are the accretion rate, and local specific energy of the TC flow, respectively. Thus the kinetic power of the emitted jet can be defined as
\begin{equation}
L_j=\eta_j\dot{B}_T=\eta_j\dot{m}_TB_T\dot{M}_{Edd}c^2=1.26\times10^{38}\eta_j\dot{m}_TB_T\frac{M_b}{M_\odot}~~~erg/s,
\label{eq1}
\end{equation}
where $M_b$ is the mass of BH in unit of $M_\odot$, $\eta_j\in[0,1]$, and $B_T>1$. The values of $B_T$ depends on the BH spin, local stochastic magnetic energy, viscosity parameter of the flow, and OBCs of the accretion flow.  If $\eta_j<0$ then may not be jet emission. 
{{The outflows/jets can be generated close to the rotation axis by the TC gas, but the detail mechanisms of the jet collimation and acceleration are still poorly understood here. However, we believe that the jet collimation and acceleration can be done by the conversion of thermal energy of TC gas and magnetic energies of both parallel flows with the help of toroidal and poloidal magnetic field in the jet. The poloidal field can be generated from the toroidal field of the disk and in jet itself, which can be due to the magneto-rotational instabilities \citep{tp96}. As we know that the magnetic field lines are highly buoyant in nature, so the toroidal field of the disk can move upward with the deflected gas of TC flow, so which can help in the collimation of the jets as illustrated in the Figure \ref{fig2}.}}

{{There are evidences in some AGNs that the base of the relativistic jets is dominated with a Poynting flux but soon becomes matter dominated with moving upward \citep[a review][]{r21}. The Blandford-Znajek (BZ) mechanism \citep{bz77} can generate the Poynting flux dominated jets and later becomes matter dominated due to the mass loading. However the mechanism of the mass loading is not much understood so far. So, we can believe that the mass loading can be done from the deflected TC gas as proposed with THAF. The deflected TC gas can reach close to the jet axis with spiraling upward, which can anchor with the magnetic field of the jet. So, the total jet power can be addition of the $L_j$ (Eq.\ref{eq1}) and the Poynting energy output due to the BZ mechanism. 
Thus the observed jet kinetic power can be greater than the accretion power with this combined matter jet generation scenario.}}
\subsubsection{Possible scenarios for existence of the KD in the accretion disk}
As we discussed that the understanding of the right path of the HID requires the TC flow. The TC flow can produce the high energy power-law components and the outflowing gas. 
The size of the KD grows with increasing accretion rate and reduces with decreasing accretion rate in the HID. However, the formation or growth of the KD flow is not clearly understood in the literatures, {such as, how does the KD maintain unique Keplerian AM distribution in large part of the disk? or Why is the outward AM transportation due to the viscosity in the disk consistent to allow only Keplerian distribution in the disk?} %
We believe that the THAF model can allow us to understand some following possible scenarios for the formation of the KD. 
\begin{enumerate}
\item    
 {It has been found that the cold-mode gas always produces the high AM accretion flows than the hot-mode inflow gas at outer AB. Because viscous AM transportation has lower efficiency in the cold-mode gas, which mostly depends as, \ie $\tau_{\mu\nu}\propto T_{AB}$, where $T_{AB}$ is the temperature of the gas at outer AB.}
So, it indicates that the cold accreting gas at AB can effectively produce the KD or KD-like flow, which can require some additional dissipative processes. 
While, the hot inflow gas at outer AB can effectively generate sub-Keplerian, hot and faster accretion flow, which can make the TC flow.  
It can be natural that the accreting gas at outer AB can have non-uniform temperature distribution. Recently, the formation of two-phase (cold gas surrounded by hot gas) inflow gas in the broad line region of the AGNs is due to a generation of thermal instability in the gas has been studied far away from the BH \citep{byz20}. So this can make the Keplerian and sub-Keplerian flows when two-phase gas accreted. {Moreover, \cite{i21a,i21b} has studied structure of an accretion ring at the outermost AB of the accretion disk. This accretion ring consists of hot thick envelop and cold thin core, which can make two-phase inflows around the BH.  So these studies are supporting our THAF scenario with two parallel flows.}

\item In THAF geometry, the whole TC flow has largely positive local specific energy, which can make flow as gravitationally unbound and possibility of the outflows/winds \citep{ny95,bb04}. We know that the outward transportation of the AM due to the viscosity is allowed to the matter move inward in the accretion flow. We believe that the same kind of the viscous processes can also work during wind outflows from the accretion flow, which can transfer AM opposite to the direction of the wind flow. So a part of AM of the upward wind flow from the TC flow can transferred toward the gas around the equatorial plane with help of the viscous process. 
Now the AM received by the gas close to the equatorial plane can condensate to form the KD or KD like flow, which can have the local thermal equilibrium (LTE).  
Recently, we have found some possibility to form the KD flow on the equatorial plane in the two-dimensional inflow-outflow study with two viscous stress tensors, $\tau_{r\phi}$ and $\tau_{\theta\phi}$ and this work is under preparation by us. {In this study, we investigated the Keplerian AM distribution on the equatorial plane. Interestingly, we found that the $\tau_{\theta\phi}$ viscous component is very effectively transferred the part of AM of upward moving gas (winds) towards the gas around the equatorial plane.}
Thus, the viscous transportation process due to $\tau_{\theta\phi}$ component can help to increase the AM of the gases around the equatorial plane, which can make maximum Keplerian AM flow. 
Moreover, effective cooling in the TC flow due to inverse-comption process can also help in the condensation of TC gas on the equatorial plane.

\item We believe that both (points 1 and 2) processes can help in formation of the KD flow. {The KD growth rate can be dependent on $T_{AB}$ of the accreting gas. We believe that if the cold accreting gas dominates over the hot gas at outer AB then the formation of the KD can be more efficient. }  
\end{enumerate}
The key point is that the both flows (hot sub-Keplerian TC, and cool KD) can be developed together with increasing the total accreting rate at AB during the right path of the HID, and 
{a high AM flow can occupy region below a low AM flow around the equatorial plane}. Interestingly, there are some observational evidences of the two simultaneous flows with different velocities in the accretion disk \citep{cgr01,shs02}.
\subsubsection{Variations of the accretion rates and state transitions}
\label{subsec:rates}
\begin{figure}
\begin{center}
\includegraphics[width=0.5\linewidth]{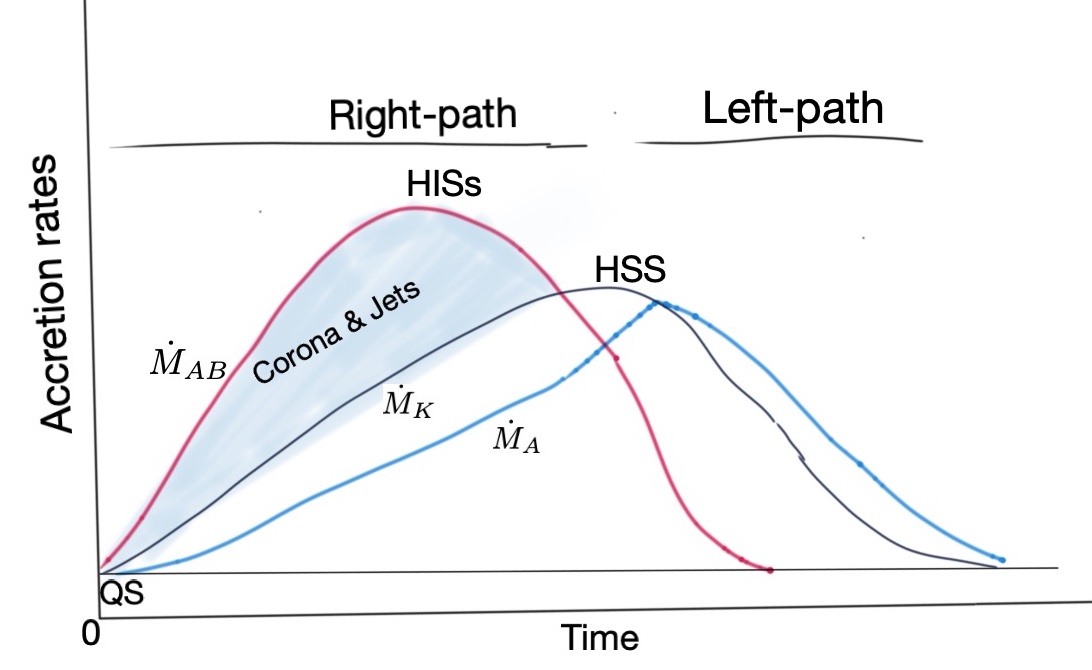}
\includegraphics[width=0.4\linewidth]{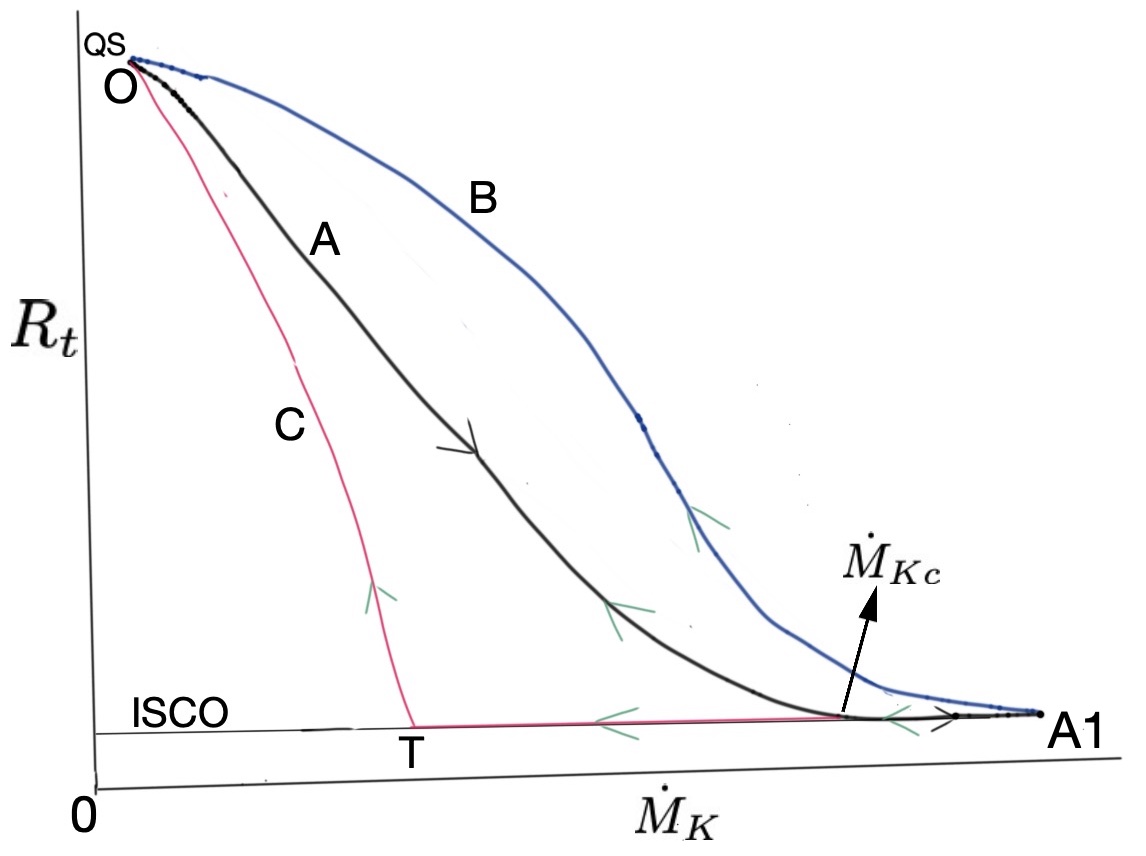}
\end{center}
 \caption{ Left panel: An illustration of a typical behavior of accretion rates with characteristic timescale for a HID with the THAF model and denoted the extreme variabilities of the spectral states. $\mab$ (red curve) is a total accretion rate at the outer AB, which includes accretion rates of the TC flow and KD flow. $\mk$ (black curve) is the accretion rate of the KD, and $\ma$ (blue curve) is an accretion rate of the ADAF or depletion/evaporation rate of the KD. The shaded region (light blue) in the $\mab$ curve can form TC flow, which can make outflows/jets and corona. Right Panel: Possible variation path of the truncation/transition radius ($\rt$) with $\mk$ for the HID.
 }
\label{fig0}
\end{figure}
{{Before interpretations of the spectral states in the BXBs is by the THAF. We need to understand the variations of the $\rt$ and the accretion rates of the disk components with following some theoretical notions and observational evidences, for instance, the lowest accretion rate corresponds to the QS, the highest accretion rate corresponds to the HSS or HIS (in case of very high IS), the generic bell-shape (similar to Gaussian or Lorentzian type profiles) variations of the accretion rates due to cyclic nature of the BXBs, $\rt$ is approaching towards BH with increasing $\mk$ during the right-path, $\rt$ is receding with decreasing $\mk$ during the left-path of the HID, and so on. 
In left panel of Figure \ref{fig0}, we have plotted three schematic curves corresponding to each component of the accretion flows in the THAF. All curves are representing only the qualitative behaviors by assuming that the lowest accretion rate corresponds to the QS, the highest of the $\mk$ curve can represent the HSS, or, the highest of $\mab$ curve can represent the HIS. The $\mab$ is the total accretion rate of the model at the outer AB, which includes the accretion rates of the TC flow and the KD means $\mab=\mk+\mt$ where $\mt$ is an accretion rate of the TC flow. $\ma$ is totally come from the $\mk$, since inner part of the KD generates the ADAF. 
Since the behavior of the BXB activities is in the cyclic nature, therefore we assume that the variation of the average accretion rate of each curve in time scale of a cycle (one HID) should be in generic bell-shape, which can be symmetric or asymmetric. Each curve may have local maxima and minima. 
We expected that any variations in the $\mk$ and $\ma$ can be dependent on the qualitative and quantitative variations in the $\mab$. We defined on the basis of the conservation of the mass flux that the $\ma\le\mk$ during right-path, and $\ma>\mk$ during left-path as shown in left-side of Figure \ref{fig0}. We believe that a presence of the TC flow ($\mt>0$) can suppress the transition region (or evaporation rate) of the KD, which can allow the $\ma<\mk$, so that $\rt$ can approach towards the BH with increasing $\mk$ (by condensation of a part of TC gas and increasing $\mab$) during the right-path. While, the $\ma>\mk$ when no or weak presence of the TC flow ($\mt\simeq0$), which can allow the receding $\rt$ with the decreasing $\mk$ (only evaporation, no condensation ($\mt\sim0$) and $\mab\simeq0$) during the left-path as shown in the curves. When $\mab$ becomes negligible at outer AB then the TC flow will quickly disappear because of high inflow velocity, but the KD flow is persistent due to very low inflow velocity, which is almost zero. So in left-path, only KD and ADAF flow are present and KD can disappear through generating the ADAF with receding $\rt$. We believed that the ADAF can generate weak power-law radiation components due to not enough interception of the soft photons with ADAF in the radial two-zone disk model and nature of the ADAF as already discussed in the subsection \ref{sec:adaf}. The presence of the TC flow can generate main power-law high radiation components and jets during the right-path. 
In the presented model, the relative size, height, and location of the peaks of the $\mab$ and $\mk$ can allow the deviations in the shape of the HIDs, which can correspond to a full outburst or failed outburst (maximum of $\mk$ can not reach to the ISCO due to weak $\mab\sim0$ so the TC flow quickly disappears). The relative heights of both peaks ($\mab$ and $\mk$) in the curves can decide the very high ISs or very HSS in the HID. Moreover, the observed peculiar variations in the BXBs and AGNs can be caused by the significant variations in the $\mab$, which can affect the accretion rates of the TC flow or KD flow or both parallel flows.}}

{{In the right panel of Figure \ref{fig0}, we have represented three symbolic curves for the variations of $\rt$ with $\mk$ by assuming that 
the inner region of KD always generates ADAF/ADAF-like flow with increasing or decreasing of $\mk$. 
It is found that the $\rt$ is the largest during the QS, corresponds to the lowest accretion rate and the smallest upto the ISCO radius during the HSS, which may corresponds to the highest accretion rate \citep{bmm11,kb15,wp21}. 
There are many paths can be possible between the ISCO radius and the largest location of $\rt$. However, only three paths are shown in right panel of the figure \ref{fig0}, which are represented between the extreme location points O (corresponds QS) and A1 (corresponds HSS). The arrows on each curve are representing the direction of path of variation of $\rt$ with changing $\mk$. Together these three curves can represent three closed/cyclic paths for studies of the HID. The cyclic paths are O$\leftrightarrow$A$\leftrightarrow$A1 (hereafter, path-A or retraced path ),  O$\rightarrow$A$\rightarrow$A1$\rightarrow$B$\rightarrow$O (hereafter, path-B) and O$\rightarrow$A$\rightarrow$A1$\rightarrow$T$\rightarrow$C$\rightarrow$O (hereafter, path-C).
The physics of the variation of $\rt$ with $\mk$ is not well explored with the theoretical as well as observational studies, for instance, a path of the $\rt$ variations with increasing or decreasing $\mk$ can be retraced same path (path-A) or different paths (path-B or -C) in a HID. Here O$\rightarrow$A1 and A1$\rightarrow$O are representing right-path with increasing $\mk$ and left-path with decreasing $\mk$ of the HID, respectively. In section \ref{sec:intro}, we have discussed the conditions for the approaching or receding or fix $\rt$ locations on the basis of conservation of mass flux in the KD, and respective relations between the $\ma$ and $\mk$. We believe that the presence of the TC flow can suppress the transition region of the KD, so which can allow $\ma\le\mk$ in right-path and otherwise $\ma>\mk$ in left-path with $\mt\sim0$. Thus the change in the relation between $\ma$ and $\mk$ in both paths of HID can be understood with presence and absence of the TC flow in our proposed THAF model. However, these relations can not be understood with two-zone disk model without some arbitrary assumptions.

Moreover, the path-A can not be possible in the studies of BXBs because it can not represent the hysteresis in the HID with using the two-zone disk model only. For understanding of the hysteresis with the two-zone model, the path of $\rt$ variation must be different with increasing and decreasing of $\mk$. A path-C type was arbitrary assumed to understand the hysteresis in the BXBs with using the two-zone disk model by \cite{kb15}. 
However, any paths between O$\leftrightarrow$A1 with increasing/decreasing $\mk$ can be possible to the explain the counter clockwise traversal and the hysteresis of HID with the THAF model, and the choice of path can affect only quantitatively shape and size of the HID, like, a full typical HID, HID of a failed outbursts, and so on. Because the presence and absence of the TC flow can naturally explain the hysteresis and counter-clockwise traversal of the HID. 

{{A $\mkc$ denotes a critical accretion rate of the KD when $\rt$ reached at the ISCO radius as represented in the right panel of Figure \ref{fig0}. This point can correspond to the HSS stage in the HID without/weak TC flow.
If the TC flow is still significantly present as a quasi-stable flow during the HSS then the inner disk can make heartbeat-like variations if the quasi-stable shocks occur in the quasi-steady TC flow, which can represent the observed heartbeat states in the GRS1915+105 \citep{fbg04}. The compression and relaxation of the inner flow due to the quasi-stable shock can make variations in the hard radiations. The quasi-periodic oscillation of a shock can happen with many combinations, like, competition between the viscous heating and cooling timescales, significance of viscous processes timescales and cooling timescales with dynamical timescale, and so on \citep[reference therein]{lck16}.
Moreover, we would like to mention that the paths A, B, and C can cross to each other depending on the variation of evaporation rate $\ma$ and growth rate of the KD via condensation of a part of TC flow or change in $\mab$. 
We would also like to mention that the variation of $\rt$ can be more complicated if the $\mt$ and $\mab$ can be changed significantly in the THAF model.
}}
\subsection{Spectral-states and transitions with THAF}\label{sec:sum}
Based on the above views with the latest theoretical understanding of the accretion flows, there are the hybrid disk configurations can possible to explain the HID, which are presented in the Fig.\ref{fig1} with assuming all the above stated possibilities and physical processes in the flow. Basically, the dominance/strength of each component of these configurations of the disk can be dependent on the quantitative and qualitative variations of the inflow gases at the outer AB, such as, a temperature of the inflow gas ($T_{vir}>>T_{AB}\sim T_{vir}$). 
The explanations of spectral states and their transitions according to the THAF are following,
\\
{\it I-QS:} It is the lowest luminosity state in the BXBs, which can have hard pawer-law with weak or negligible BB component. So the QS can be represented by the disk geometry of panel  (A) of Figure \ref{fig1}, which can have weak ADAF and weak KD with low $\mab$. Here the ADAF and TC may not be much distinguishable due to their low accretion rates with $\rt>10^3\rs$ \citep{wp21}. 
The QS stage (Fig.\ref{fig1}A) has the lowest accretion rate. When the accretion flow is evolved enough with the increasing compactness of gas in the space around the BH with increasing accretion rate 
 then a BXB enters to a LHS stage. 
\\ 
{\it II-LHS:} It is more luminosity than the QS with higher $\mab$. In this stage, the disk becomes stronger than the QS as disk geometry is shown in Figure \ref{fig1}B. In LHS, the spectrum is dominated in the high energy power-law radiation, which can be originated from the hot TC flow. 
The Figure \ref{fig1}B shows three flows, which are denoted as TC (blue, solid line), KD (magenta, dash-dotted line) and ADAF/ADAF-like (red, dotted line) with the accreting gas at the AB, \ie $\mab$, and the arrows represent the directions of the gas motion.
The TC flow can be stronger and hotter. The hot TC flow can generate high energy radiations via Inverse-Comptonization process, Bremsstrahlung and synchrotron radiative processes. 
$\rt$ can be usually $\lsim10^2\rs$ in the LHS. The jets can be originated from the TC flow due to upward deflection of the TC gas via the strong surface of higher momentum flux of the ADAF as discussed in subsection \ref{subsec:jet}. The TC flow can have steady/quasi-steady shocks in the inner region of the flow. The post-shock region can produce the quasi-periodic oscillations in the non-thermal radiations with the shock variations \cite{scl20,mkc13,ddp20,soa21}.
\\
{\it III-HISs:} Here $\mab$ can be reached to highest value as shown in the left panel of Figure \ref{fig0}. We believe that the source can enter to HIS stage when $\rt$ is around $10-20\rs$ or less, and disk structure can be looked like Figure \ref{fig1}C. This stage can be divided into two parts, one can be a hard-HIS and other can be a soft-HIS. The hard-HIS can correspond to a rising part of upper $\mab$ curve, and the soft-HIS can correspond to a decreasing part of the $\mab$ curve. In hard-HIS, the jet become stronger than the jet in the LHS due to increase in the strengths of the TC flow and ADAF with increasing the accretion rates of the both and decreasing $\rt$. 
In the soft-HIS stage, the KD starts to dominate over the hot TC flow due to the decreasing $\mt$ and the photon from the KD can significantly cool down the TC gas by inverse-Compton process in the inner part of the TC flow with increasing $\mk$. So, the both decreasing $\mt$ and cooling dominating process can help to collapse the TC flow close to the BH, so the jet generation process can also be stopped, and the source enters to the HSS stage. The ADAF can also be collapsed if $\rt$ reached to the ISCO radius.
\\ 
{\it IV-HSS:} In this stage, the $\mk$ and $\ma$ can be reached to maxima of the curves (Figure \ref{fig0}) and $\mab\sim0$ or $\mt\sim0$. The disk structure can be represented as in the Figure \ref{fig1}D. For a typical HID, we believed that the $\rt$ can reach to the ISCO radius and corresponding $\mk$ can be denoted as $\mkc$. So, the disk structure will be the KD dominated, which can make the soft spectrum dominated with multicolor blackbody radiation. If a maximum of the $\mk$ curve of a HID cycle is occurred before reaching to the $\mkc$ or ISCO radius due to lack of the $\mab$ then we can call this situation as a failed outburst, which does not follow the HSS stage after the outburst or collapse of the jet due to $\mt\sim0$. However, the ADAF can be presented here because $\rt$ location is away from the ISCO radius.
\\
{\it V-LIS:} This stage can have only the two-zone disk model with large $\rt>10\rs$ with $\mk<\mkc$ and $\mab\sim0$. The generated ADAF can have $\ma\ge\mk$ with a large transition region in the inner part of the KD due to lack of the TC flow. The disk structure is represented in Figure \ref{fig1}E. In this stage, the soft photon decreases with the increasing $\rt$ (means moving away from the BH) with decreasing $\mk$, but the contribution part of the hard photons increases with large ADAF size due to large $\rt$. When the $\mk$ can become very weak then the source enters to the QS stage.

The possible roles of the ADAF disk in our model can be i) produce weak intrinsic radiations (Bremsstrahlung and Synchrotron emission), and soft-excess (Compton thick scattering){, which depends on the $\ma$ and size of the ADAF, \ie $\rt$}, ii) make barrier (centrifugal and pressure supported surface) against the in-falling gas of TC, 
{ so the TC gas can make the jet close to the rotation axis due to deflection from the ADAF surface}, and  iii) The preceding gas of TC flow can be compressed due to the ADAF barrier, and if it is compressed enough then the TC flow can make strong shock. Although, the shocks can be generated without the additional ADAF barrier as shown in many theoretical and numerical studies around different BHs \citep{kc17,kg18,ky21,lck16,gbc20,soa21,smd22}, and the post-shock region is much hotter than other flows, which can produce very high energy Comptonized radiations, such as, hard $X-$ray, and soft $\gamma-$ray.
The generated shocks can be steady or quasi-steady, which can make quasi-periodic variations in the emitted high energy radiations \cite{smg22}.

 Moreover, the two-parallel flows in the THAF model can have possibilities of many physical processes, such as, magnetic reconnection, exchange of matter and energy via AM transportation, viscous heating and cooling, and the jet generation, evolution and launching during the right-path of the HID.
 Recently, it has been found in the MHD simulation that the magnetic field (both toroidal and poloidal) is rapidly evolved in the post-shock region of the hot flow than in other part of the disk, which can be helpful to make outflows faster and collimated \citep{gbc20}. 
 So, the shocked TC flow is suitable for the magnetic field (MF) evolution and jet generation. In THAF model, if the ADAF can thread with the MF lines. So these MF lines can be hold for longer time against the gas of TC flow. Since the ADAF is always slow moving inward than the TC flow. 
 Therefore, the possibility of the interaction of MF lines of both parallel hybrid flows can be higher than the co-moving of the MF lines in the same flow. Thus this situation can be suitable for the magnetic reconnection with the interception of the MF lines of fast TC flow and slow ADAF in the inner region as shown in the Fig.\ref{fig2}. 
\begin{figure*}
\begin{center}
\includegraphics[width=0.3\linewidth]{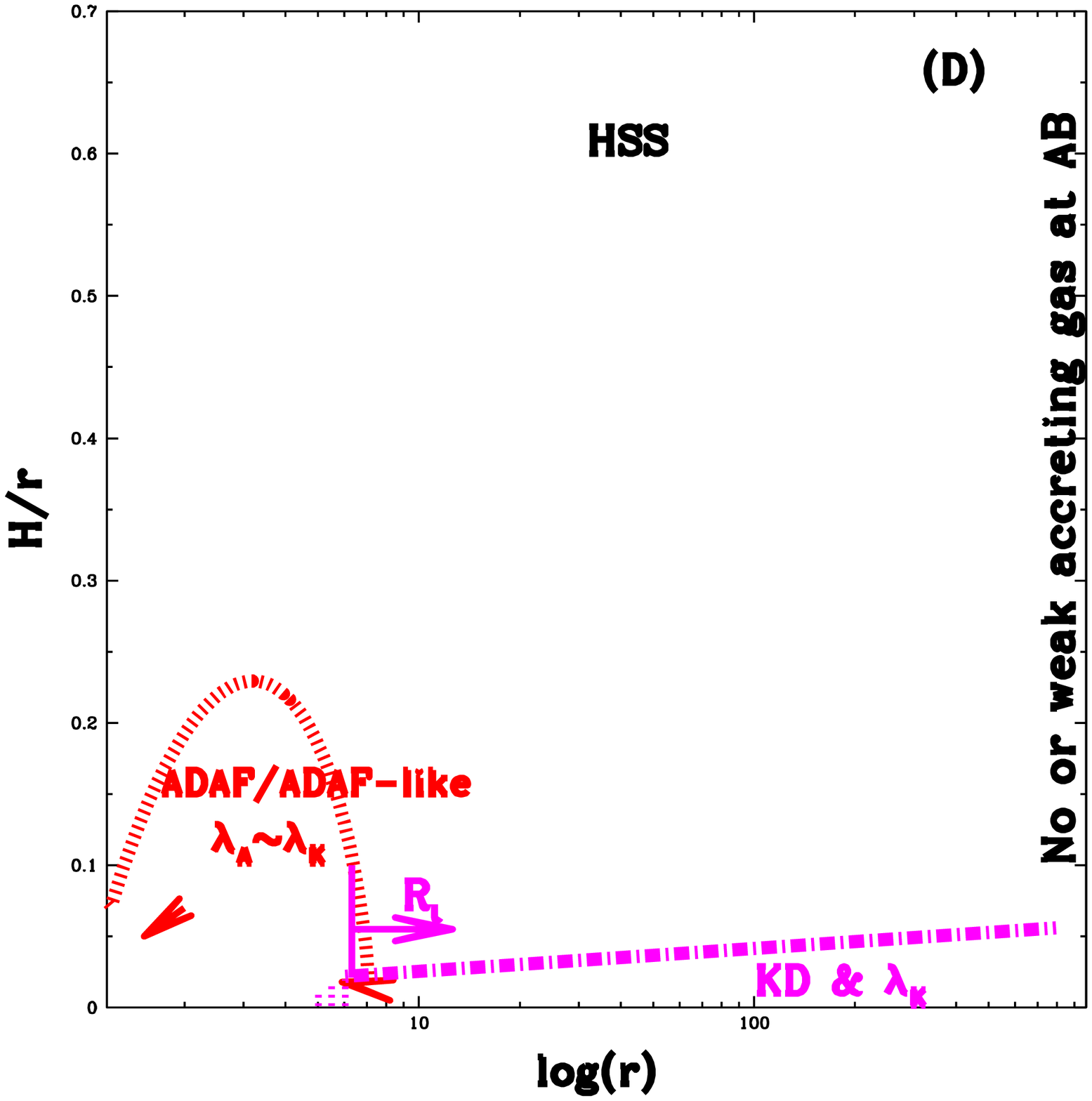}
\includegraphics[width=0.3\linewidth]{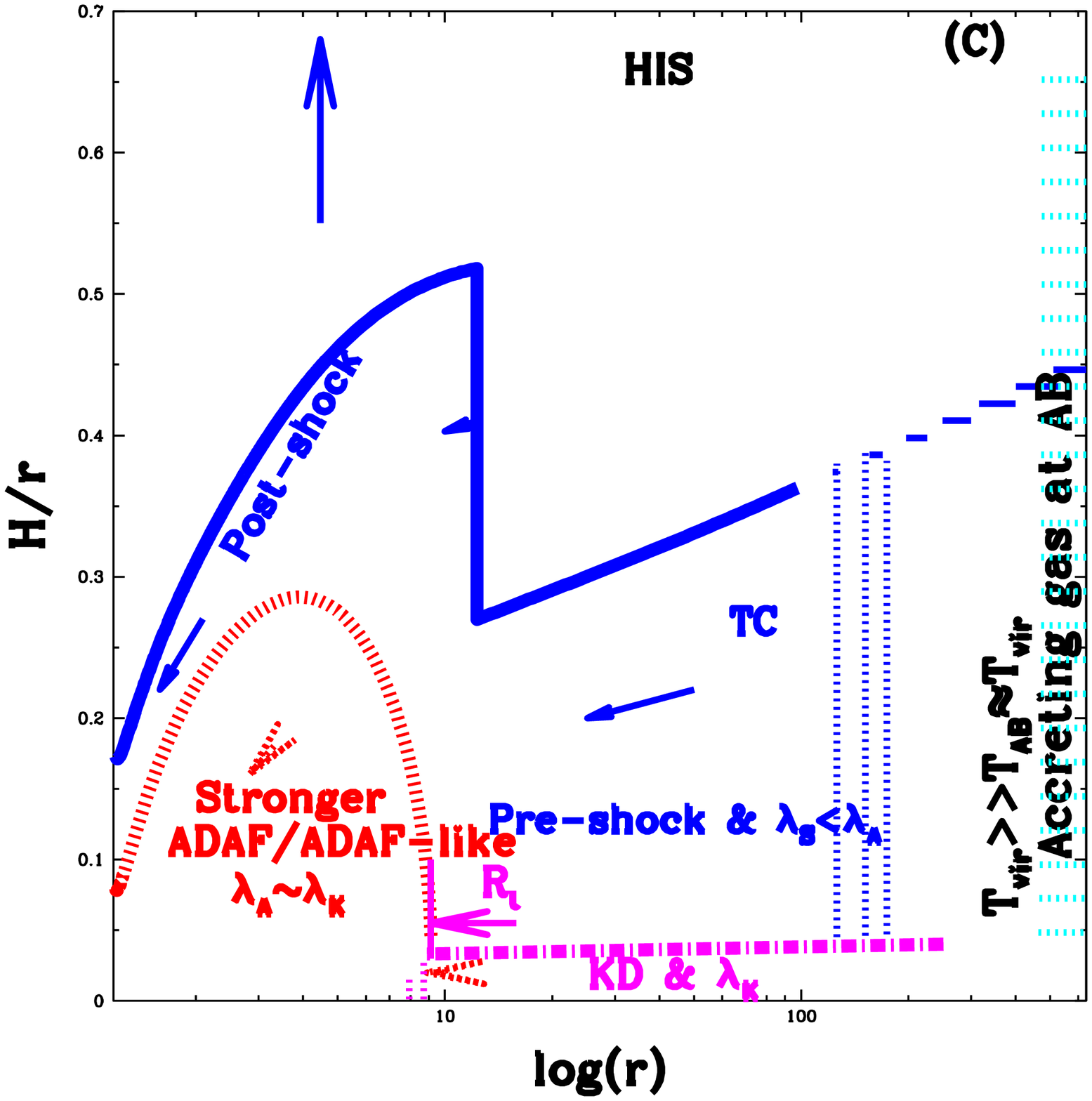}
\end{center}
\begin{center}
\includegraphics[width=0.2\linewidth]{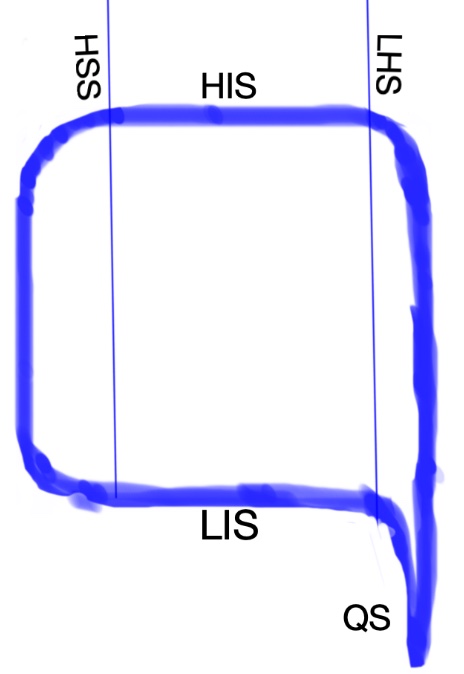}
\includegraphics[width=0.3\linewidth]{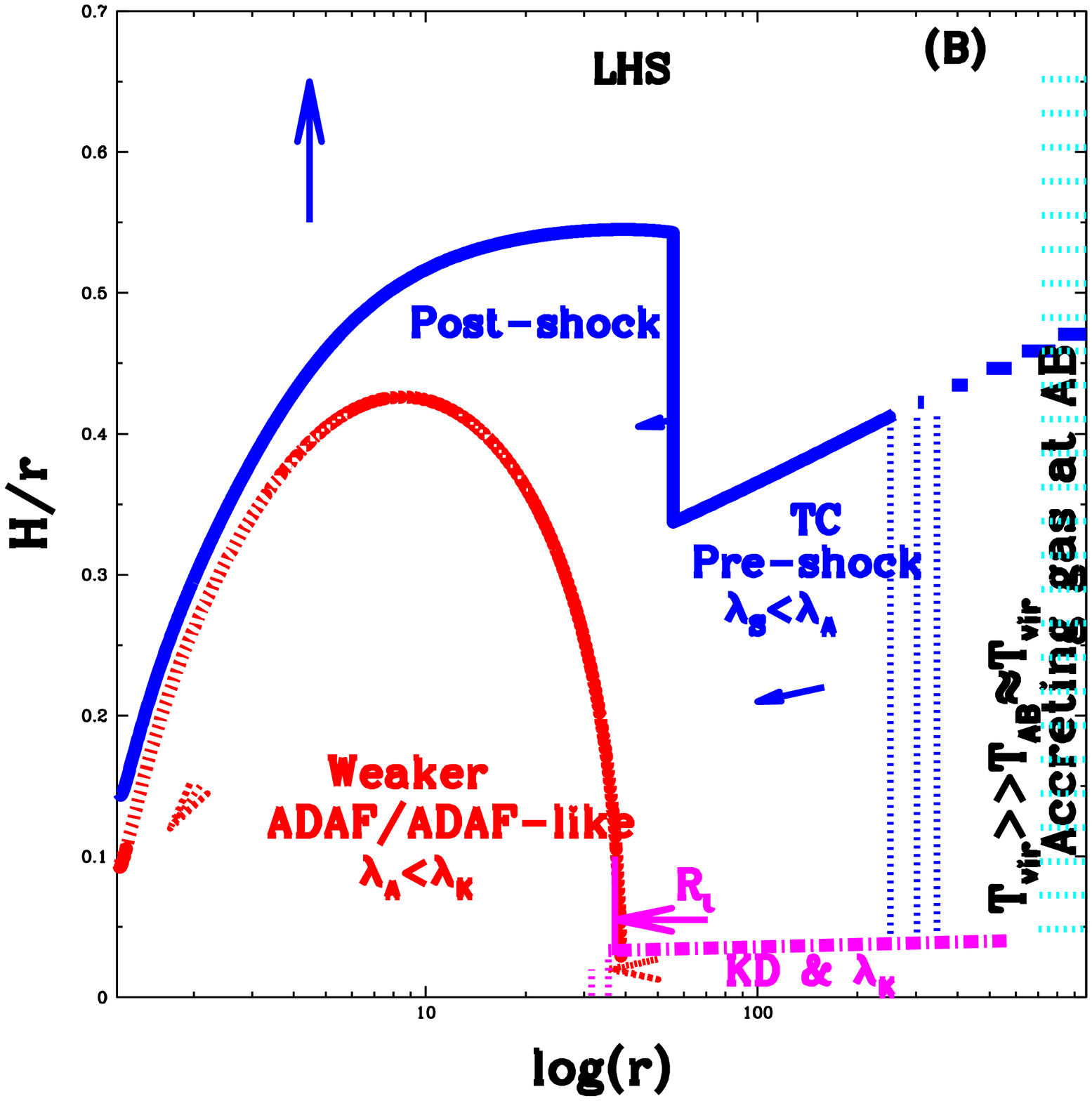}
\end{center}
\begin{center}
\includegraphics[width=0.3\linewidth]{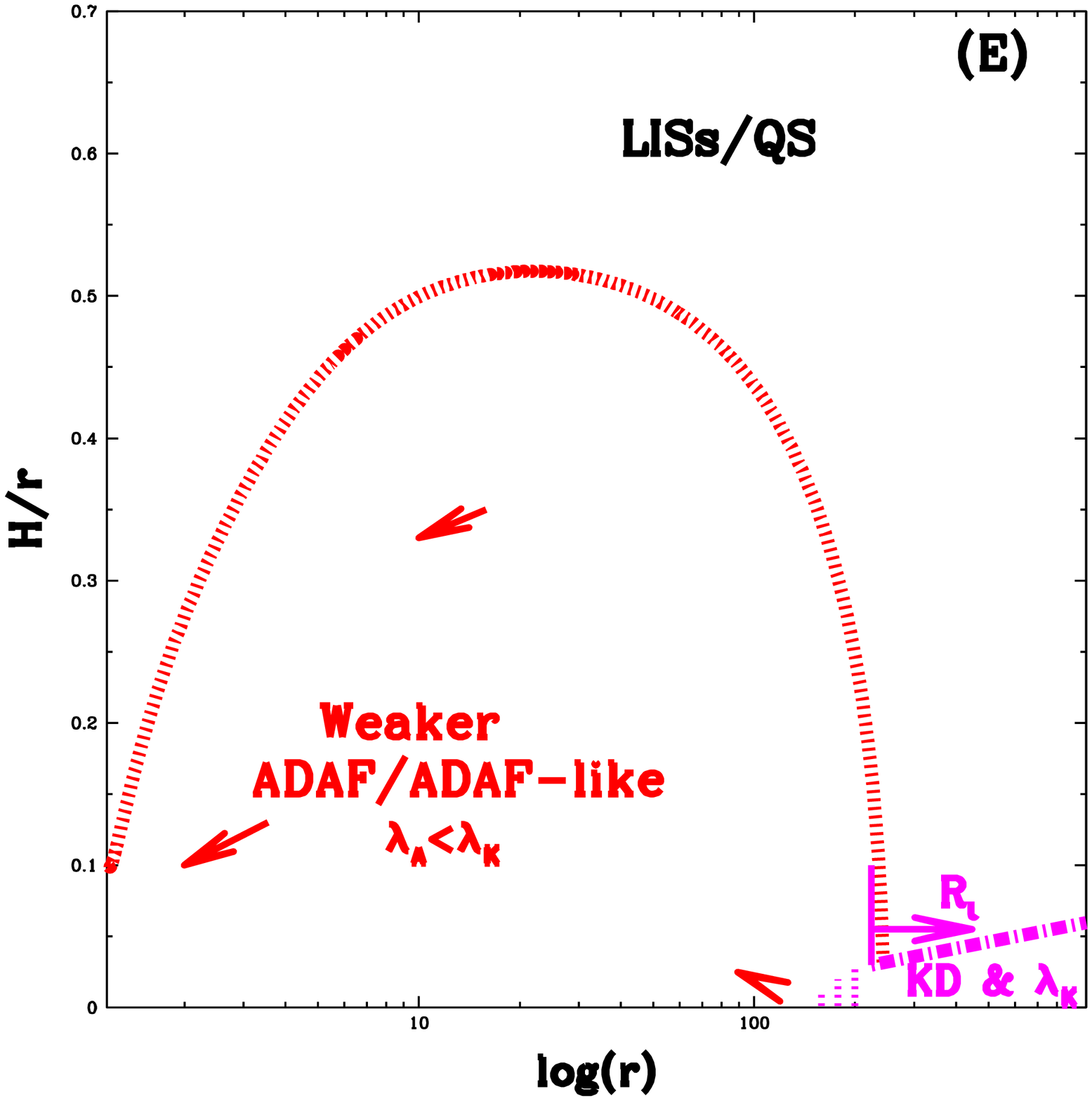}
\includegraphics[width=0.3\linewidth]{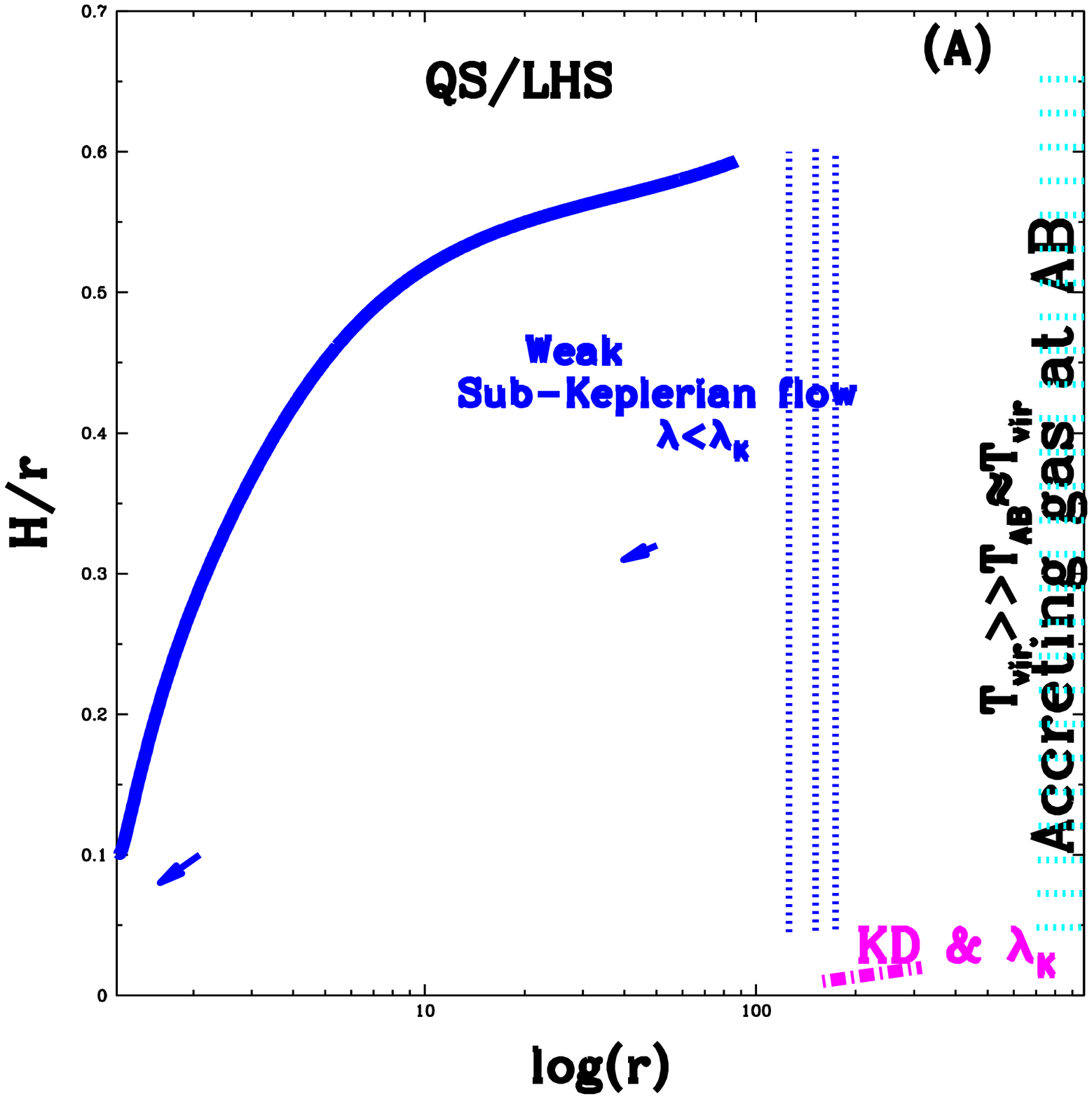}
\end{center}
\caption{
An illustration of the q$-$shape curve of the HID in middle of first column, and typical disk structures from A to E, which can represent the various spectral states of BXBs. The $H/r$ height profiles of the accretion flows are obtained from the numerically solving general relativistic viscous fluid equations around a Kerr BH, except the KD flow (dash-dotted line). 
The vertical line (magenta color) is representing transition radius ($R_t$), and the arrows are representing gas moving direction. The vertical jump in the TC flow represents shock location for a particular model parameters. $T_{\rm AB}$ and $T_{\rm vir}$ are the local temperature of inflow gas and virial temperature at the outermost AB, respectively. 
\label{fig1}}
\end{figure*}
\begin{figure}
\begin{center}
\includegraphics[width=\linewidth]{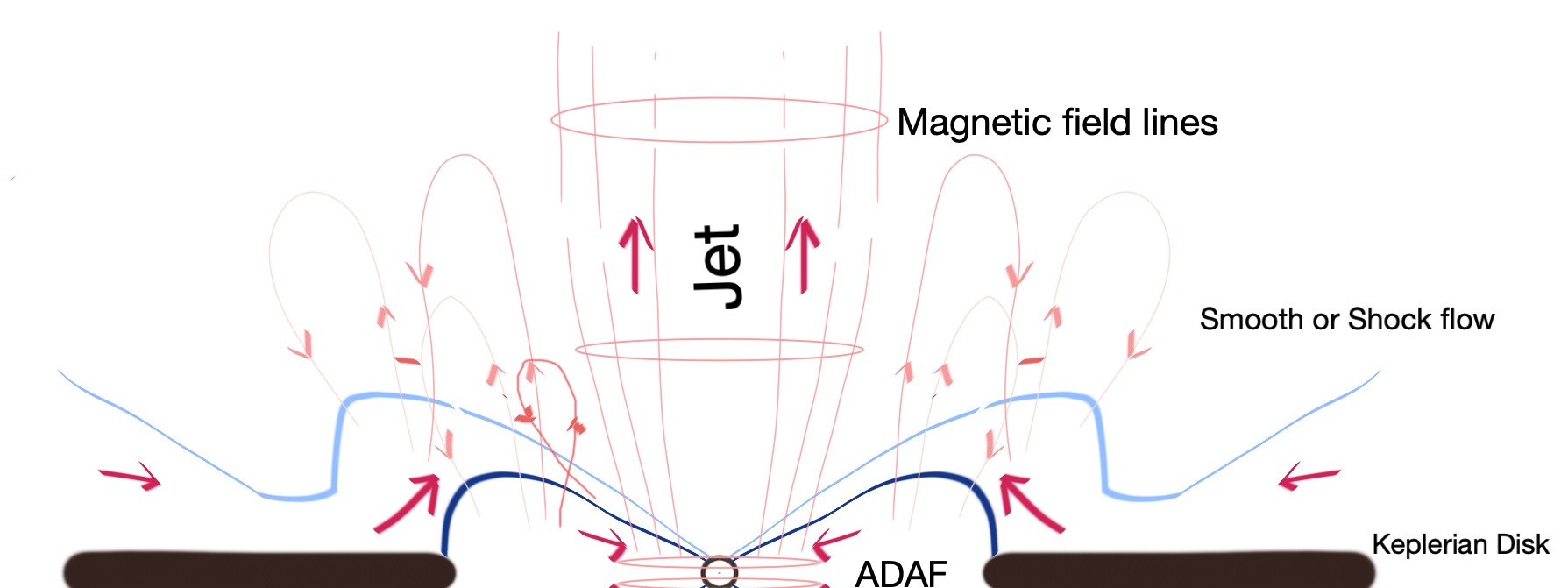}
\end{center}
 \caption{
A schematic illustration of the disk-jet system for the proposed THAF model.
 }
\label{fig2}
\end{figure}
\section{Conclusions}\label{sec:con}
We have addressed the inconsistencies of the two-zone disk model with theoretical as well as observational points of views, and tried to resolve those inconsistencies with including the TC flow in it.
{{The proposed THAF model can naturally explain the presence of the hysteresis and the counter clockwise traversal of the q- curve in the HID with increasing accretion rate ($\mab$), which can make the two parallel hybrid flows, TC and KD flow. The TC flow is hotter, which can act as a corona and also make the jets/outflows during the right-path. The TC is absent during the left-path of the HID when $\mab\sim0$, which makes automatically different paths in the HID. 
The TC flow is sub-Keplerian and faster moving. So it can quickly change the intensities or luminosities of the source due to intermittent changes in the flow than the KD flow. The intermittent changes can be at outer AB (qualitative and quantitative changes in the inflow gas) or within the disk due to formation of steady/quasi-steady shocks or gas feedback due to failed winds or change in the magnetization level of the flow.
 %
These intermittent changes can help to understand the peculiar properties of the sources, for instance, fast changes in the AGNs means CLAGNs, heartbeat states in the BXBs, and some peculiar quasi-periodic variabilities (by quasi-stable shocks in the TC flow). Moreover, the TC can also change the accretion rate of the KD flow via the condensation of the gas when the succeeding gas gains the AM during the upward motion of the TC gas as discussed in the section \ref{sec:3cf}. In the presented model, it is much suitable to understand the conditions for the approaching of the $\rt$ with increasing $\mk$ during the right-path and receding of the $\rt$ with decreasing $\mk$ during the left-path of the HID with presence and absence of the TC flow. The presence of the TC flow can suppress the transition region of the KD, which can allow $\ma<\mk$, and $\ma>\mk$ if the TC flow is absent as discussed in the subsection \ref{subsec:rates}.}} 

The two parallel hybrid flows can generate possibilities of many physical processes, such as, exchange of matter and energies via condensation, viscous processes (heating and AM transportation) and radiative processes, magnetic reconnections, jet generation and so on.
{{The surface of the ADAF can work as the partial/strong barrier for the TC flow. This barrier helps to reduce the BH gravity pull for the inflowing TC gas with the deflecting and driving the TC gas upward as the outflows/jets. These outflows can be accelerated and collimated by the conversion of the thermal energy and magnetic energy with the help of the poloidal and toroidal fields. Moreover, %
it is known from the study of the particle dynamics around the central object that the particle with Keplerian AM has most stable circular orbit, and a stable particle is highly resistive to any perturbations.  
We found that the inner ADAFs are always having higher AM distribution than the advective TC flow \citep{ky21}. The configurations of the KD and inner ADAF dominate in the left-path of the HID.
Thus, the disk configuration of the left-path of the HID can be more stable than the disk configuration of the right-path with any intrinsic or extrinsic perturbations in the THAF model. So we can conclude from the  theoretical point of view that the right-path can have high variability nature than the left-path of the HID, which is also indicated in the observations of the BXBs \citep{bmm11}. 
}}  

However, we need to understand the detailed physics of the THAF in the future due to the presence of the two parallel accretion components in the disk, such as, to understand the interaction of two parallel advective flows (ADAF and TC) in the inner part of the disk and two parallel the KD and TC in the outer part of the disk, and explore the formation of the KD with the TC flow with the presence of the different nature of accreting gases at the AB. These things can be explored in our future studies via theoretical and simulation investigations with including relevant physical processes in the fluid flow. 
We believe that a suitable numerical setup can generate the two parallel flows of the hot sub-Keplerian and cool Keplerian disk flows due to the different viscous and radiative time scales in the different nature of inflowing gases at the AB.
Moreover, we also need to explore the detail variations of $\rt$ with $\mk$ in the observational studies of the full outbursts and failed outbursts of the BXBs .
So the path of variation  of the transition radius can help us to provide more precise theoretical model(s) with proper physics. We are hoping that the coming future facility XRISM can help us to find the more relevant physical model by the studies of accretion physics at high $X$-ray resolution \cite{gkd22}.
We are also expecting some modeling studies of the BXBs with the THAF model in the near future.
\section*{Acknowledgement}
This work is funded by the National Natural Science Foundation of China (grant No. 11725312) and National SKA Program of China No. 2020SKA0120300. R. K. is supported by CAS President's International Fellowship Initiative (PIFI), Grant No. 2021PM0017. 
R.K. also thanks the Harish-Chandra Research Institute (HRI), Prayagraj, India for the local hospitality and stay during the preparation of this manuscript.
\bibliographystyle{unsrtnat}
\bibliography{THAF}

\end{document}